\documentclass[structabstract]{aa}
\usepackage{natbib} 
\usepackage{graphicx}
\usepackage{makecell}
\usepackage{txfonts}
\begin{document}

   \title{High redshift galaxies in the ALHAMBRA survey }
   
   \subtitle{I. selection method and number counts based on redshift PDFs\thanks{Based on observations collected at the German-Spanish Astronomical Center, Calar Alto, jointly operated by the Max-Planck-Institut f\"ur Astronomie (MPIA) at Heidelberg and the Instituto de Astrof\'{\i}sica de Andaluc\'{\i}a (CSIC)}}

   \titlerunning{High redshift galaxies in the ALHAMBRA survey}

   \author{K. Viironen
          \inst{1}
          \and
          A. Mar\'{i}n-Franch
          \inst{1}
          \and
          C. L\'opez-Sanjuan
          \inst{1}
          \and
          J. Varela
          \inst{1}
          \and
          J. Chaves-Montero
          \inst{1}
          \and
          D. Crist\'obal-Hornillos
          \inst{1}
          \and
          A. Molino
          \inst{2,3}
          \and
          A. Fern\'andez-Soto
          \inst{4,5}
          \and
          G. Vilella-Rojo
          \inst{1}
          \and
          B. Ascaso
          \inst{6,3}
          \and
          A. J. Cenarro
          \inst{1}
          \and
          M. Cervi\~no
          \inst{3,7,8}
          \and
          J. Cepa
          \inst{7,8}
          \and
          A. Ederoclite
          \inst{1}
          \and
          I. M\'arquez
          \inst{3}
          \and
          J. Masegosa
          \inst{3}
          \and
          M. Moles
          \inst{1,3}
          \and
          I. Oteo
          \inst{9,10}
          \and 
          M. Povi\'c
          \inst{3}
          \and
          J. A. L. Aguerri
          \inst{7,8}
          \and
          E. Alfaro
          \inst{3}
          \and
          T. Aparicio-Villegas
          \inst{11,3}
          \and
          N. Ben\'itez
          \inst{3}
          \and
          T. Broadhurst
          \inst{12}
          \and
          J. Cabrera-Ca\~no
          \inst{13}
          \and
          J. F. Castander
          \inst{14}
          \and
          A. Del Olmo
          \inst{3}
          \and
          R. M. Gonz\'alez Delgado
          \inst{3}
          \and
          C. Husillos
          \inst{3}
          \and
          L. Infante
          \inst{15}
          \and
          V. J. Mart\'inez
          \inst{5,16}
          \and
          J. Perea
          \inst{3}
          \and
          F. Prada
          \inst{3}
          \and
          J. M. Quintana
          \inst{3}
          }
   \institute{Centro de Estudios de F\'{i}sica del Cosmos de Arag\'on, Plaza San Juan 1, planta 2, 44001 Teruel, Spain\\
             \email{kerttu@cefca.es}
             \and
             Instituto de Astronomia, Geof\'isica e Ci\^encias Atmosf\'ericas, Universidade de S\~ao Paulo, S\~ao Paulo, Brazil
             \and
             Instituto de Astrof\'isica de Andaluc\'ia (IAA-CSIC), Glorieta de la astronom\'ia s/n, 18008 Granada, Spain
             \and
             Instituto de F\'{i}sica de Cantabria, Avenida de los Castros s/n, 39005 Santander, Spain
             \and
             Unidad Asociada Observatori Astronomic (IFCA-UV), C/ Catedr\'atico Jos\'e Beltr\'an 2, 46980 Paterna, Spain
             \and
             GEPI, Paris Observatory, 77 av. Denfert Rochereau, 75014 Paris, France
             \and
             Instituto de Astrof\'{i}sica de Canarias, V\'{i}a L\'actea s/n, La Laguna, 38200 Tenerife, Spain
             \and
             Departamento de Astrof\'{i}sica, Facultad de F\'{i}sica, Universidad de la Laguna, 38200 La Laguna, Spain
             \and
             Institute for Astronomy, University of Edinburgh, Royal Observatory, Blackford Hill, Edinburgh EH9 3HJ
             \and
             European Southern Observatory, Karl-Schwarzschild-Str. 2, 85748 Garching, Germany
             \and
             Observat\'orio Nacional, COAA, Rua General Jos\'e Cristino 77, 20921-400 Rio de Janeiro, Brazil
             \and
             Department of Theoretical Physics, University of the Basque Country UPV/EHU, Bilbao, Spain
             \and
             Departamento de F\'isica At\'omica, Molecular y Nuclear, Facultad de F\'isica, Universidad de Sevilla, Spain
             \and
             Institut de Ci\`encies de l'Espai (ICE-CSIC), Facultat de Ci\`encies, Campus UAB, 08193 Bellaterra, Spain
             \and
             Departamento de Astronom\'ia, Ponticia Universidad Cat\'olica. Santiago, Chile
             \and
             Departament d'Astronomia i Astrof\'iısica, Universitat de Val\`encia, 46100 Burjassot, Spain
}
   \date{}

 
  \abstract
   {Most observational results on the high redshift restframe UV-bright galaxies are based on samples pinpointed using the so-called dropout technique or Ly-$\alpha$ selection. However, the availability of multifilter data now allows the dropout selections to be replaced by direct methods based on photometric redshifts. In this paper we present the methodology to select and study the population of high redshift galaxies in the ALHAMBRA survey data.
}
   {Our aim is to develop a less biased methodology than the traditional dropout technique to study the high redshift galaxies in ALHAMBRA and other multifilter data. Thanks to the wide area ALHAMBRA covers, we especially aim at contributing to the study of the brightest, least frequent, high redshift galaxies.}
   {The methodology is based on redshift probability distribution functions (zPDFs). It is shown how a clean galaxy sample can be obtained by selecting the galaxies with high integrated probability of being within a given redshift interval. However, reaching both a complete and clean sample with this method is challenging. Hence, a method to derive statistical properties by summing the zPDFs of all the galaxies in the redshift bin of interest is introduced.}
   {Using this methodology we derive the galaxy rest frame UV number counts in five redshift bins centred at $z=2.5, 3.0, 3.5, 4.0$, and 4.5, being complete up to the limiting magnitude at $m_{\rm{UV}}$(AB)=24, where $m_{\rm{UV}}$ refers to the first ALHAMBRA filter redwards of the Ly-$\alpha$ line. With the wide field ALHAMBRA data we especially contribute to the study of the brightest ends of these counts, accurately sampling  the surface densities down to $m_{\rm{UV}}$(AB)=21-22.}
   {We show that using the zPDFs it is easy to select a very clean sample of high redshift galaxies. We also show that it is better to do statistical analysis of the properties of galaxies    using a probabilistic approach, which takes into account both the incompleteness and contamination issues in a natural way.}


   \maketitle

\section{Introduction}

Identifying and studying high redshift galaxies is crucial for our understanding of the early epochs of galaxy evolution. At the beginning of the nineties, the implementation of the so-called dropout technique opened the era for detections of copious numbers of these early galaxies \citep[e.g.][]{guhathakurta90,steidel92,steidel93,steidel96a,steidel96b}. These galaxies are discovered based on their  broadband colours, i.e. by measuring the drop in brightness due to the Lyman break at rest frame 912~\AA~ and/or the Lyman forest between 912~\AA~and 1216~\AA. For high redshift galaxies ($z\geq$ 2) these features are detected at optical or infrared wavelengths and permit the detection of these so-called Lyman-break galaxies (LBGs) from the ground. The dropout technique is sensitive to galaxies that are young enough to produce copious amounts of ultraviolet light, and are sufficiently dust free for a fair amount of this light to escape the galaxy.

Detections of high redshift galaxies opened the possibility for observational studies of some fundamental questions of galaxy evolution and cosmology at early epochs. One of the most widely studied properties are the LBG rest frame ultraviolet (UV) luminosity functions. The UV luminosities of the galaxies (once corrected for dust extinction) are directly proportional to their star formation rates. Hence, the study of the UV luminosity density, derived by integrating the luminosity function at different redshifts, gives information about the star formation history in the Universe.

Lyman-break galaxies can also act as tracers of dark matter at high redshift through the study of their clustering properties. The formation history of galaxies is basically understood through two fundamental evolutionary processes, i.e. the production of stars and the accumulation of dark matter. While the baryonic matter, i.e. stars, gas, and dust, can be studied through the light they emit, the dark matter cannot be directly detected using  electromagnetic waves. However, the clustering properties of galaxies are closely related to the distribution and amount of the underlying dark matter \citep[see][and references therein]{ouchi04a}.

Most of these studies, up to the very recent ones, have applied the dropout technique for candidate selection \citep[e.g.][and many more]{ouchi04a,ouchi04b,shim07,reddy08,ly11,bouwens14}. While this technique is efficient at selecting high redshift galaxies, it is also affected by significant incompleteness and contamination, losing some fraction of the population at the selected redshift, or allowing galaxies at other redshifts to enter the sample. While the latter can be dealt with by obtaining spectroscopic redshifts \citep[see e.g.][]{steidel96a,steidel96b,reddy06}, the former remains a serious difficulty. We are not yet at the point of spectroscopic blind surveys, hence, a step forward towards less biased candidate selection is offered by multifilter surveys. They combine the efficiency and unbiased nature of photometric surveys with very low resolution spectral information, permitting us to derive more information on the surveyed objects such as their accurate photometric redshifts.

Many authors \citep[e.g.][]{shim07,ly11} have combined the data of their colour selected LBG samples with information at other passbands in order to carry out spectral energy distribution (SED) fitting and to  derive more information on the objects in question, like their photometric redshifts. However, basing the actual candidate selection on photometric redshifts as, for example, \citet{mclure06} have done, has only recently started to become a common practice. As discussed by \citet{mclure11}, when multifilter data are available this approach has several advantages over the traditional colour selection. It makes the best use of the available information in multiple filters, it should be less biased as any colour preselection is not required, and it directly offers the photometric redshifts for the galaxies of interest and allows the competing photometric redshift solutions at low redshift to be investigated. Recently, \citet{lefevre14} have used photometric redshifts to select an unbiased target sample of high redshift galaxies for the VUDS survey. Photometric redshift selection is also used in the framework of the CLASH survey \citep[e.g.][]{bradley14} and in the recent works of \citet{finkelstein14} and \citet{bowler14}.

In this paper we introduce a method for studying high redshift ($z\sim 2-5$) galaxies based on their photometric redshifts. Our study makes use of the complete redshift probability distribution functions (zPDFs), rather than  the best redshift (i.e. the median derived from the zPDF or the highest peak of the zPDF). We show how a very clean candidate selection can be made based on the zPDFs and discuss how this technique also suffers from contamination issues if completeness is tried to be reached. Finally, we discuss why for many statistical purposes candidate selection is not needed. Instead, these studies can be based directly on the redshift (and the corresponding luminosity, mass, star formation rate, etc.) probability distributions. As an example we present probabilistic number counts for several high redshift bins. These counts should be free from contamination and incompleteness issues, if the used zPDFs correctly reflect the uncertainties in the redshift estimations.

The method is developed and tested with the data from the Advanced Large, Homogeneous Area Medium Band Redshift Astronomical \citep[ALHAMBRA,][]{moles08} Survey. The total area used for our study is 2.38 deg$^2$, covered with 20 medium band optical filters, plus $J, H$, and $K_s$ in the near infrared (NIR). In addition to the novel methodology, an advantage of our ALHAMBRA high redshift galaxy study as compared to the previous LBG studies is the large area the survey covers, split into eight independent fields, reducing biases due to the cosmic variance and allowing the study of the rarest, brightest, high redshift galaxies. Galaxies at $z\sim 1$ in ALHAMBRA (plus GALEX, IRAC, MIPS, and PACS) were studied in \citet{oteo13a,oteo13b}. In this first paper about ALHAMBRA $z > 2$ galaxies, we concentrate on the methodology of studying the galaxy properties using the whole information in their zPDFs. In the subsequent papers, this methodology will be applied to studying the properties of these galaxies.

The methodology presented here is generic and can be applied to any multifilter data set with accurate zPDFs, such as the data from the SHARDS \citep[the Survey for High-z Absorption Red and Dead Sources,][]{perez_gonzalez13}, and the future J-PLUS (Javalambre Photometric Local Universe Survey, Cenarro et al., in prep.) and J-PAS \citep[Javalambre-PAU Astrophysical Survey,][]{benitez14b}.

The outline of this paper is as follows. Section \ref{sec:data} describes the ALHAMBRA data used for this study. Section \ref{sec:photoz} gives an introduction to the ALHAMBRA photo-z derivation and its validity for high redshift galaxies. In Sect. 4 the sample selection is described, the contamination and completeness of the sample are discussed, and our sample selection is compared with the traditional dropout selections. In Sect. \ref{sec:prob} the probabilistic approach is introduced and the rest frame UV number counts are derived. A summary is given in Sect. \ref{sec:sum}. Where necessary, we assume $\Omega_{\rm{m}}$ = 0.3, $\Omega_{\Lambda}$= 0.7, and H$_0$ = 70 km s$^{-1}$ Mpc$^{-1}$. Magnitudes are given in the AB system \citep{oke83}.

\begin{figure*}
\centering
\includegraphics[width=\linewidth]{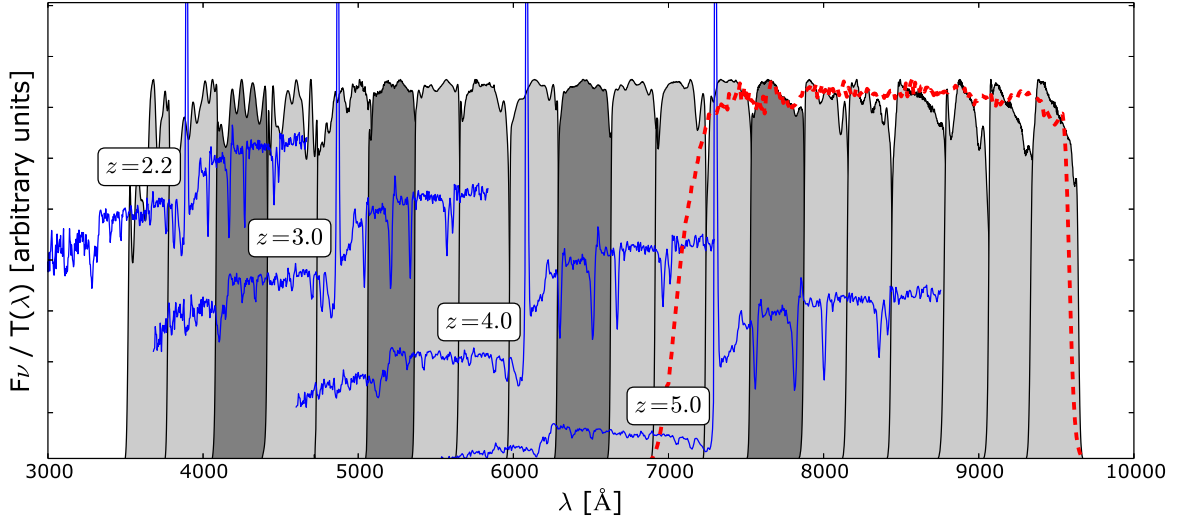}
\caption{The $z\sim3$ composite LBG spectrum of \citet{shapley03} moved to redshifts $z=2.2, 3.0, 4.0,$ and 5.0 considering the variation of the intergalactic opacity with redshift (blue lines). For clarity, the spectra are shifted vertically and plotted only up to 1460\AA~(restframe). The ALHAMBRA optical filter transmission curves are overplotted as shaded grey areas, and the dashed red line corresponds to the synthetic $F814W$ filter. The first filter redwards of the Ly-$\alpha$ line in each redshift is marked in darker grey. [{\it A colour version of this figure is available in the online edition.}]}
\label{fig:filts}
\end{figure*}

\section{Data}\label{sec:data}

ALHAMBRA \citep{moles08} has mapped a total of 4 deg$^2$ of the northern sky in eight separate fields during a seven year period ($2005-2012$). Of the total surveyed area, 2.8 deg$^2$ have been completed with all the filters (2.38 deg$^2$ after masking, as will be detailed in Sect.~\ref{sec:sel}). ALHAMBRA uses a specially designed filter system \citep{aparicio-villegas12} that covers the optical range from 3500~\AA~to 9700~\AA~with 20 contiguous, equal width ($\sim$300~\AA~FWHM), medium band filters, plus the three standard broadbands, $J$, $H$, and $K_s$, in the NIR. The photometric system has been specifically designed to optimise photometric redshift depth and accuracy \citep{benitez09}. The observations were carried out with the Calar Alto 3.5m telescope using two wide field cameras: LAICA in the optical, and OMEGA-2000 in the NIR. The 5$\sigma$ limiting magnitude reaches $\gtrsim24$ for all filters below 8000 \AA~and increases steeply towards redder medium-band filters, up to m(AB) $\sim 21.5$ for the reddest optical filter at 9700~\AA~\citep[see Fig. 37 of][]{molino14}. In the NIR the limiting magnitude is $\sim 23$ for $J$, $\sim 22.5$ for $H$, and $\sim 22$ for $K_s$. For details about the NIR data reduction see \citet{cristobal-hornillos09}, while the optical reduction is  described in Crist\'obal-Hornillos et al. (in prep). The ALHAMBRA object catalogues and the associated Bayesian photometric redshifts (BPZs) are described in \citet{molino14} and are available through the ALHAMBRA web page\footnote{http://alhambrasurvey.com}. At the moment only the best BPZs are public; the full zPDFs will be published in the future. In Fig.~\ref{fig:filts} we show the transmission curves of the optical ALHAMBRA filters together with the $z\sim3$ composite spectrum of 811 LBGs of \citet{shapley03} moved to different redshifts.

\section{Photometric redshifts}\label{sec:photoz}

The work in this paper relies on the photometric redshifts provided for all the objects in the ALHAMBRA catalogue as detailed by \citet{molino14}. These photometric redshifts were estimated using BPZ2.0 (Ben{\'i}tez et al., in prep), an updated version of the Bayesian Photometric Redshift (BPZ) code \citep{benitez00}. This code uses Bayesian inference where a maximum likelihood, resulting from a $\chi ^2$ minimisation between the observed and predicted colours for a galaxy among a range of redshifts and templates, is weighted by a prior probability. The maximum likelihood (ML) method may suffer from colour--redshift degeneracies (like 4000\AA~break vs. Lyman break) and the inclusion of a suitable prior information can help to break these degeneracies. However, both maximum likelihood  and Bayesian redshift probability distributions are available for all the ALHAMBRA sources.

The BPZ2.0 SED library \citep[see][]{molino14} consists of 11 SEDs: five templates for elliptical galaxies, two for spiral galaxies, and four for starburst galaxies along with emission lines and dust extinction. The opacity of the intergalactic medium has been applied as described in \citet{madau95}. The prior used gives the probability of a galaxy with apparent magnitude $m_0$ having a certain redshift $z$ and spectral type $T$. The prior has been empirically derived for each spectral type and magnitude by fitting luminosity functions provided by GOODS-MUSIC \citep{santini09}, COSMOS \citep{scoville07}, and UDF \citep{coe06}.

For  each catalogued ALHAMBRA object both the maximum likelihood and Bayesian redshift probability distribution functions (zPDFs) are given separately for each template used in the $\chi ^2$-fitting. We are not interested in limiting ourselves to any galaxy type, hence, we use the redshift PDFs integrated over all templates, and normalised to one:

\begin{equation}\label{eq:pint1}
\int PDF(z) {\rm d}z=\int \int PDF(z,T) {\rm d}z {\rm d}T = 1.
\end{equation}

These zPDFs give the probability along the redshift axis of a galaxy in question to be at that redshift. Hence, the probability, $p$, that a galaxy is within the redshift bin $z_1 < z < z_2$ is

\begin{equation}\label{eq:pint2}
p=\int_{z_1}^{z_2} \! PDF(z) {\rm d}z.
\end{equation}

\subsection{ALHAMBRA redshifts for high$-z$ galaxies}\label{sec:alhz}

The first questions to solve before blindly using the photometric redshift information for analysing high redshift galaxies are: Can we really trust these redshifts for high-z galaxies? Is it more reliable to use the maximum likelihood (ML) or the Bayesian, full probability (FP), redshift probability distributions?

The idea of the prior information is to reduce the redshift estimation uncertainties. However, the prior information should be used only if it really can be trusted. The complete census of high redshift galaxies is still poorly known and the known census is most probably biased \citep[see e.g.][]{lefevre14}. Hence, using any prior information based on such a census could introduce undesired biases or uncertainties. For this reason, our answer to the second question above would a priori be to base our study on the ML redshift information. This will be further studied in the following.

\begin{figure*}
\centering
\includegraphics[width=\linewidth]{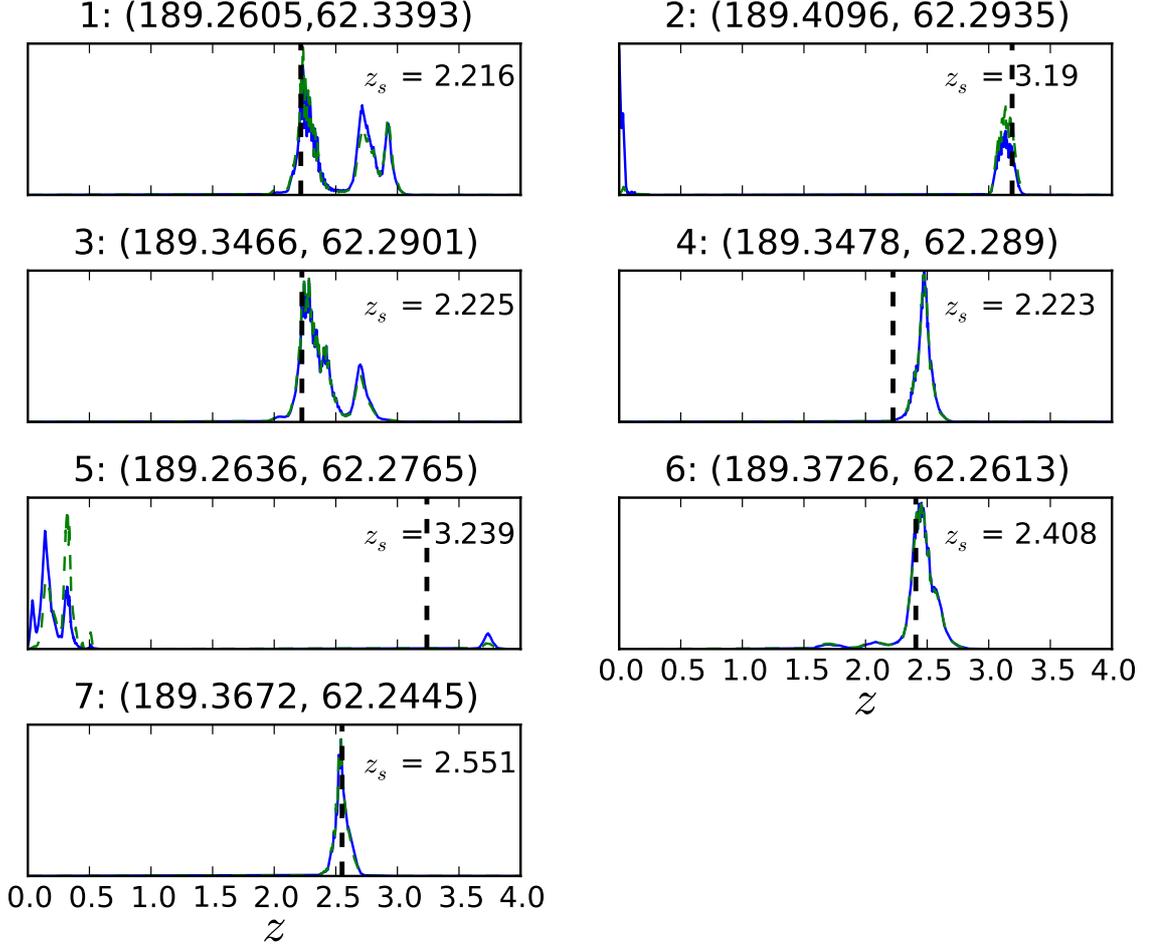}
\caption{The maximum likelihood (solid blue line) and Bayesian (dashed green line) zPDFs for eight galaxies with spectroscopic redshifts. The spectroscopic redshifts ($z_s$) are given in each panel and are also marked as dashed black vertical lines. See the text for more details. [{\it A colour version of this figure is available in the online edition.}]}
\label{fig:specz}
\end{figure*}

While the accuracy of the ALHAMBRA BPZs is well tested and demonstrated \citep{molino14} for galaxies up to z$\sim 1.5$ (being $\sim 1\%$), this is not the case for the galaxies that we are interested in because of  the small number of spectroscopic redshifts for high redshift galaxies in ALHAMBRA area. In the sample of $\sim 7200$ galaxies with spectroscopic redshifts used to verify the ALHAMBRA photo-z accuracy \citep{molino14}, there are only 57 with redshifts above $z=2.2$ (the lowest redshift at which the Lyman forest would be sampled by at least  one ALHAMBRA filter). Of these only 12 are brighter than $m=24$ in the first filter redwards of the Ly-$\alpha$ line. A literature search reveales that five of these are classified as quasars, and as the BPZ template library does not include quasar spectra, we do not expect to be able to accurately recover  their redshifts. Hence, we are left with seven spectroscopically confirmed bright normal high redshift galaxies. In Fig.~\ref{fig:specz} we show the ALHAMBRA ML and FP zPDFs for these seven galaxies together with their ALHAMBRA coordinates and spectroscopic redshift \citep[from][]{barger08}. We see that for five of them (Nos. $1$, $3$, $4$, $6$, $7$), the redshift is reasonably well recovered, $\Delta z\lesssim 0.3$, where $\Delta z$ is the difference between the first peak of the zPDF and the spectroscopic redshift. For galaxy $4$, whose shift between the spectroscopic and photomnetric redshift is the largest of the five, the shift corresponds almost exactly to a width of one filter, i.e. it seems BPZ has  mistaken the location of the Ly-$\alpha$ break by one filter. Of the remaining two, for galaxy $2$, the first peaks of both ML and FP zPDFs are located at low redshift, but the peaks at high redshift enclose most of the probability. Galaxy $5$ shows a secondary peak at high redshift (higher than the spectroscopic redshift), but most of the probability resides at low redshift. The spectra of these objects are not public making it hard to further study the reason for these discrepancies between the spectroscopic and photometric redshifts. However from the ALHAMBRA SEDs we infer that, most probably, these discrepancies derive from the common confusion between the 4000\AA~break and Lyman break.

To have a better control on the expected redshifts, and a wider range of magnitudes to be tested, we carried out a simulation. For this purpose we used the $z=3$ composite LBG spectrum of \citet{shapley03}. We moved this spectrum to different redshifts: $z = 2.2, 3.0, 4.0, 5.0$. The lowest redshift was selected such that the Lyman forest would be sampled at least by one ALHAMBRA filter, while considering the previous work on LBG number counts \citep[e.g.][]{yoshida06}, we do not expect to discover many galaxies above $z=5$ owing to the magnitude limits of ALHAMBRA.

To simulate the different redshifts, the original spectrum was first moved to redshift z=0 removing the effect of cosmic opacity using the equations of \citet{madau95}, then the same equations were used to simulate the spectra at different redshifts. The original spectrum cover the wavelength range from 920 \AA~to 2000 \AA. To cover the whole ALHAMBRA optical wavelength range in the simulated redshifts, we artificially extended it assuming a flat behaviour of the UV continuum ($F_{\nu}=$constant, i.e. $F_{\lambda}\propto\lambda^{-2}$) from 2000 \AA~redwards up to the Balmer break at 4000 \AA, and from 920 \AA~bluewards, down to 912 \AA~where the flux is assumed to drop abruptly adopting a cosmic opacity $\tau_{eff}=10$ for $\lambda <912$ \AA. 

The resulting spectra were convolved with the ALHAMBRA filters. The convolved spectra were scaled to the desired magnitudes (at the first filter redwards from the Ly-$\alpha$) to sample the magnitude range $m = 20.2 - 24.0$. The lower magnitude was defined so that we really could expect to have galaxies of this magnitude at our lowest redshift bin \citep[see][]{ly11}, while the upper limit was set to reach the ALHAMBRA sensitivity limit.

We also considered realistic errors for each magnitude at each filter. To obtain these, we selected one arbitrary ALHAMBRA field and studied how the magnitude error varied with magnitude for each filter. Using all the objects in the field, we created {\it mag} vs. {\it mag\_err} curves for each filter and found the best fitting solutions of the form $mag\_err = a + b*e^{c*mag}$. The expected errors at each magnitude and filter were then obtained from these equations and assigned to the simulated LBG spectra.

Finally, a Monte Carlo simulation was carried out. Each LBG spectrum was perturbed inside its error bars 100 times. When the simulated magnitude was below the 1$\sigma$ detection limit (adapted again from one arbitrary ALHAMBRA field), it was replaced by this 1$\sigma$ limiting magnitude, as required by BPZ for non-detections. The BPZ code was run for each of the simulated spectra to obtain both their FP and ML redshift probability distributions. We studied the recovered distributions with two questions in mind: 1) How well can we  recover LBGs as high redshift galaxies? For this, we used equation (\ref{eq:pint2}) and tested how often the galaxies would be recovered to have a probability $p>0.9$ to be within a redshift bin $1.9 < z < 5.3$. The redshift bin was selected to be wider than the range of the input redshifts so that small errors in redshift would not place the borderline objects outside the tested range; and 2) How accurately is the redshift of these simulated galaxies  recovered?

The recovery rate of LBGs as high-redshift galaxies is summarised in Table~\ref{tab:zdist}, where the percentage of simulated LBGs having a probability greater than 90\% to be within the desired redshift range for each input redshift and magnitude are listed for both the FP and ML redshift distributions. We see that, in general, the recovered fraction is worse for the $z=2.2$ LBGs than for the higher redshift sources. We assume that the lower redshifts are recovered with less accuracy, because the lower the redshift, the less pronounced is the characteristic Ly-$\alpha$ break and it is seen with fewer filters. We also see that while the ML method recovers the high redshift nature of the simulated galaxies very well, the Bayesian approach gives worse results. It systematically fails for the brightest magnitudes, reducing the probability of the LBGs to be at high redshift below our 90\% limit, and also starts failing for the fainter magnitudes earlier than the ML approach. 

We note that according to earlier studies \citep[see][]{yoshida06}  for galaxies at redshift $z\geq4$ we possibly could not expect to observe rest frame UV magnitudes brighter than 22. This would partially justify why the Bayesian approach fails to recover the redshifts of these galaxies. If in addition the brightest end of the \citet{ly11} surface density plots were  dominated by interlopers, the redshift recovered by the Bayesian method could actually be reasonable. However, knowledge of the high redshift galaxy population is still very incomplete and most probably biased. Hence, basing any study on prior knowledge of such a population might be dangerous and could lead to further biases as  our simulation also indicates. We do not want to take the risk of losing the especially interesting bright objects. For these reasons, we decided to base our study on the ML redshift probability distributions, i.e. to assume a flat prior.

\begin{table}
\caption[]{The percentage fraction of simulated LBGs of different redshifts and magnitudes fulfilling our selection criterion.}
\label{tab:zdist}
\centering
\begin{tabular}{c|c c c c|c c c c}
\hline
& \multicolumn{4}{c}{Maximum likelihood} & \multicolumn{4}{c}{Bayesian}\\
\hline\hline
\diaghead{\theadfont z}%
{ $z$}{Mag }&
\thead{2.2}&\thead{3.0}&\thead{4.0}&\thead{5.0}&\thead{2.2}&\thead{3.0}&\thead{4.0}&\thead{5.0}\\
& & & & & & & &\\
\hline
20.2 & 100 & 100 & 100 & 100 & 0   & 0   & 0   & 0 \\
21.  & 99  & 100 & 100 & 100 & 0   & 0   & 0   & 0 \\
22.  & 95  & 100 & 100 & 100 & 70  & 14  & 9   & 1 \\
23.  & 73  & 100 & 100 & 100 & 61  & 99  & 100 & 0 \\
24.  & 55  & 95  & 99  & 97  & 38  & 68  & 77  & 35 \\
\hline
\end{tabular}
\end{table}

\begin{figure}
\centering
\includegraphics[width=0.9\linewidth]{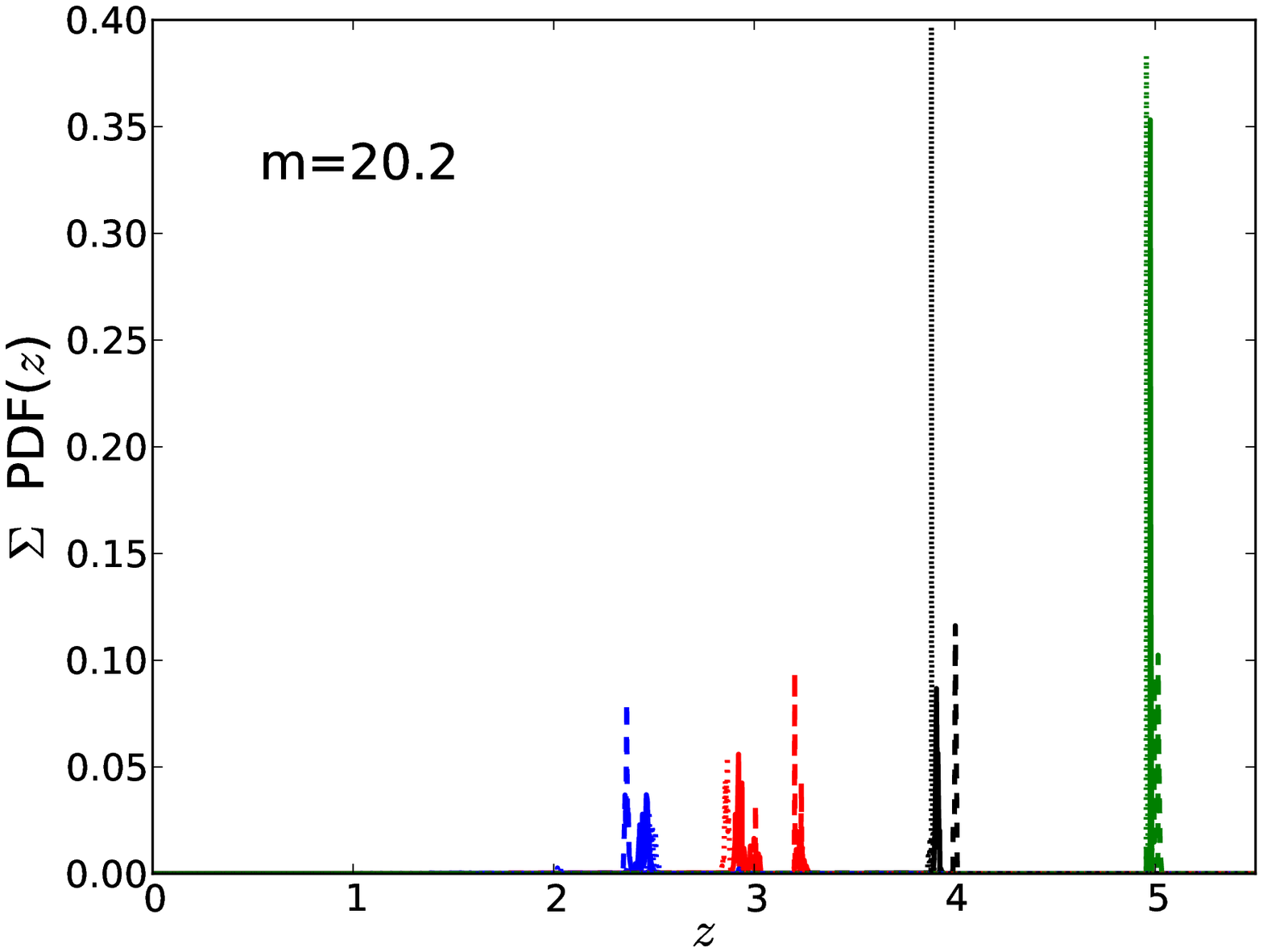}
\includegraphics[width=0.9\linewidth]{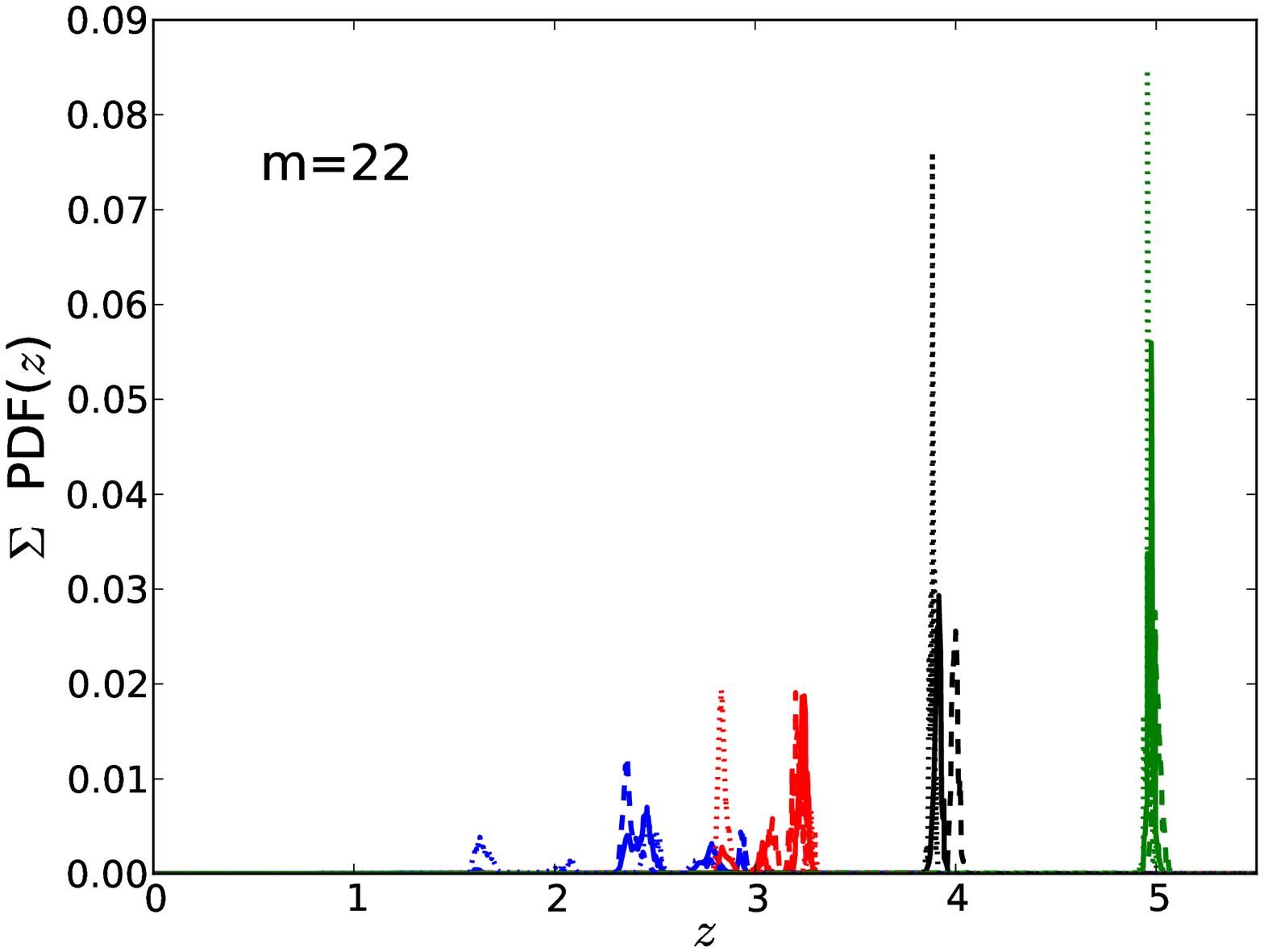}
\includegraphics[width=0.9\linewidth]{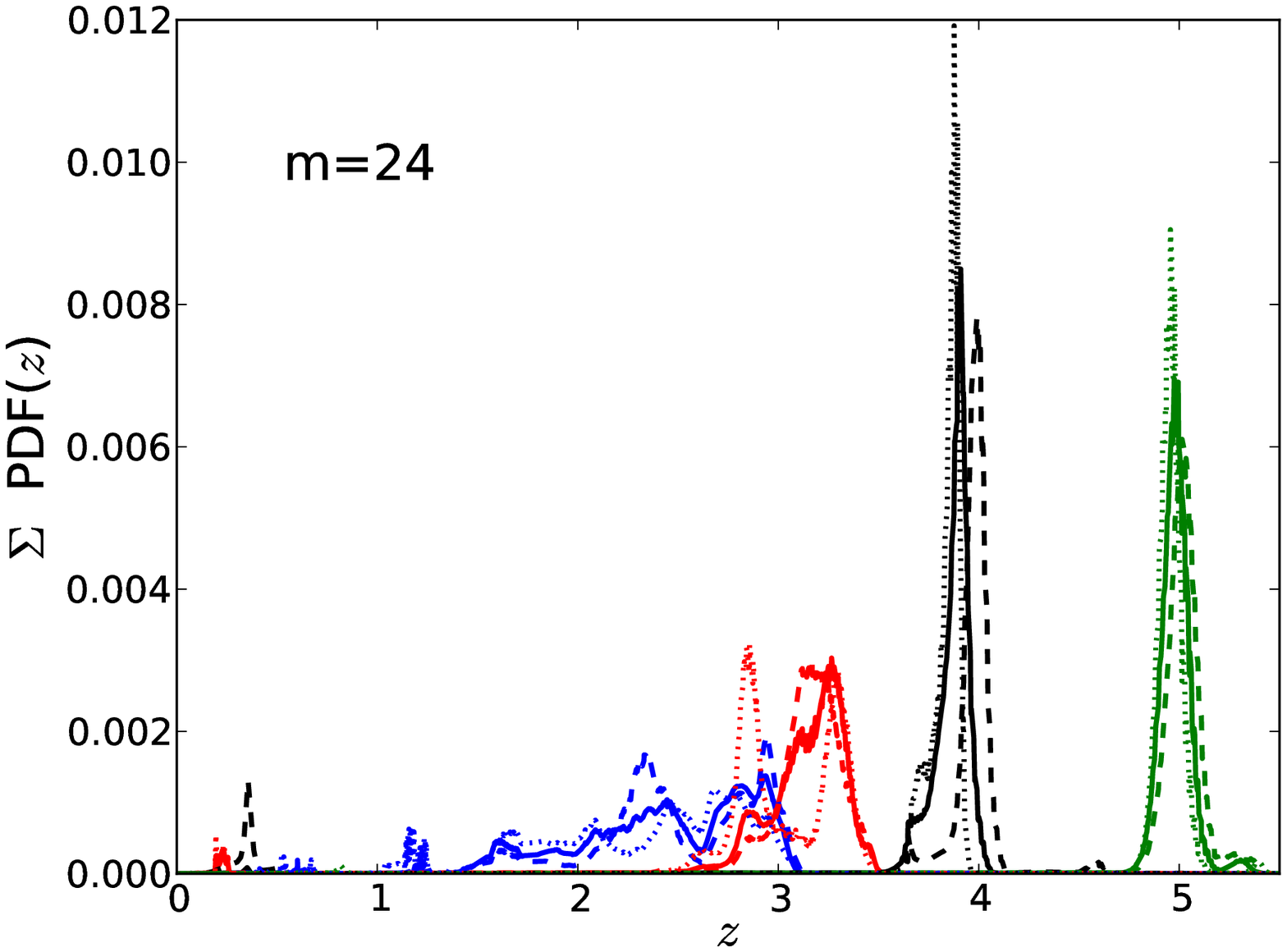}
\caption{Recovered summed redshift distributions, normalised to one at the integrated probability, for 100 simulated LBGs at z=2.2 (blue lines), z=3.0 (red lines), z=4.0 (black lines), and z=5.0 (green lines) for three different rest frame UV magnitudes. The solid lines correspond to the simulation with the original composite LBG spectrum, and the dashed and dotted lines to the simulations with the same spectrum, but the Ly-$\alpha$ line removed and doubled, respectively. [{\it A colour version of this figure is available in the online edition.}]}
\label{fig:hists}
\end{figure}

\begin{table}
\caption[]{The recovered redshifts for 100 simulated LBGs at different magnitudes and redshifts. Presented are the average value and its standard deviation derived from Gaussian approximations of the summed zPDFs in Fig.~\ref{fig:hists}.}
\label{tab:z}
\centering
\begin{tabular}{c c c c c}
\hline
Mag & $z\_{\rm in}$=2.2 & $z\_{\rm in}$=3.0 & $z\_{\rm in}$=4.0 & $z\_{\rm in}$=5.0\\
\hline\hline
20.2 & 2.46 $\pm0.02$ & 2.92 $\pm0.02$ & 3.909 $\pm0.007$ & 4.975 $\pm0.001$\\
21.0 & 2.44 $\pm0.04$ & 3.2 $\pm0.2$ & 3.91 $\pm0.01$ & 4.975 $\pm0.008$\\
22.0 & 2.5 $\pm0.2$ & 3.2 $\pm0.1$ & 3.91 $\pm0.01$ & 4.97 $\pm0.02$\\
23.0 & 2.5 $\pm0.3$ & 3.2 $\pm0.1$ & 3.90 $\pm0.03$ & 4.97 $\pm0.03$\\
24.0 & 2.5 $\pm0.5$ & 3.2 $\pm0.2$ & 3.89 $\pm0.07$ & 4.98 $\pm0.06$\\
\hline
\end{tabular}
\end{table}

To test the accuracy of the recovered redshifts, we summed the ML zPDFs of the simulated LBGs in order to see how well the input redshifts were recovered. In Fig.~\ref{fig:hists} we show these summed zPDFs at three different magnitudes. In Table~\ref{tab:z} we list the input redshifts and the recovered average redshifts and their sigma, derived from Gaussian approximations of the summed zPDFs. The recovered redshifts generally show a bias towards smaller or higher $z$  than the input redshift, the bias becoming smaller with increasing $z$. It is not surprising that the redshift is worse recovered at the lower simulated $z$, as at lower $z$ the Lyman forest is sampled by fewer ALHAMBRA filters (see Fig.~\ref{fig:filts}), and the Ly-$\alpha$ break is less pronounced at lower redshifts. However, it is intriguing to see that even though the recovered redshift becomes more and more peaked towards higher $z$, a systematic bias towards smaller redshift remains. Bayesian Photometric Redshift templates do not include the Ly-$\alpha$ emission line, while this line is present in the composite spectrum used for the simulations. The presence of the line could dilute the Ly-$\alpha$ break and cause the bias in the redshift estimation towards lower $z$. To test this hypothesis, we manually removed the Ly-$\alpha$ line from the composite spectrum, and repeated the simulation. We also repeated the simulation doubling the Ly-$\alpha$ line strength. We have plotted the resulting summed zPDFs in Fig.~\ref{fig:hists} together with the original results. In the two largest simulated redshifts the tendency of increasing Ly-$\alpha$ line strength to increasingly underestimate the redshift is obvious. At the two lower redshifts this is not enough to explain the involved uncertainties. However, in all the simulated redshifts the average sizes of the biases are $\Delta z \leq 0.3$, and, since we work with rather rough redshift bins, we consider the obtained accuracy acceptable.

\section{A sample selection approach}\label{sec:sel}

In this section we present one way of selecting a clean sample of high redshift galaxy candidates, using the zPDFs, and check it against traditional dropout selections. Our sample selection consists of two steps: cleaning the catalogue from non-desired detections, and applying a redshift selection. While the first step is always needed, we will discuss later that, while sometimes useful, for many purposes a redshift selection is actually not needed.

\subsection{Catalogue selection}\label{sec:catsel}

We start our candidate selection by cleaning the ALHAMBRA catalogues of any possible spurious or false detections, duplicated detections, and stars. For this purpose we used the masks defined in \citet{arnalte14} describing the sky area which has been reliably observed, and the stellar flag provided in the ALHAMBRA catalogues \cite[see][]{molino14}, setting "Stellar\_Flag" $<0.51$ in order to remove stars. This should remove the stars up to $m<22.5$ in the reference filter, {\it F814W}. Above this magnitude the stellar flag is not defined, and slight contamination by faint stars may remain. However, for fainter magnitudes, the fraction of stars  compared to galaxies declines rapidly, with a contribution of $\sim10$\% for magnitudes $m(F814W) = 22.5$, declining to $\sim 1$\% for magnitudes $m(F814W) = 23.5$ \citep{molino14}. After these steps, our data consist of a total of 362788 galaxies in 2.38 deg$^2$.

\subsection{The redshift selection}

There is no one single correct way of applying zPDFs for candidate selection. The best redshift (e.g. the first peak) can be derived from the zPDF and assigned to each galaxy \citep[e.g.][]{lefevre14} or the zPDF can be integrated and used in one way or another to select a list of candidates \citep[e.g.][]{mclure11,duncan14}. Here we use the second approach in a very simplified way in order to select a clean sample of high redshift ALHAMBRA galaxies. With this approach one is not obliged to be limited to any specific redshift range. However, we limit our study to the redshift range $2.2 < z \leq 5.0$. The lower limit is set so that we sample  the Lyman forest, i.e. the spectrum bluewards of the Ly$\alpha$ line, with at least  one filter. Because of the depth of ALHAMBRA we do not expect to find many galaxies at the upper limit of $z>5.0$. In addition, the ALHAMBRA sensitivity limit worsens rapidly for wavelengths above $\sim 8000$\AA, and with the upper redshift limit we make sure to measure the UV continuum redwards the Ly$\alpha$ break in at least  two filters bluer than 8000\AA.

When all of the information on the redshift probability distribution is used, one can select as candidates all the galaxies that have  a probability greater than a given threshold of being at the desired redshift interval. This threshold can then be selected to obtain the desired balance between completeness and contamination. To introduce this technique, we decided to opt for a clean selection and select as candidates the objects fulfilling the criterion

\begin{equation}\label{eq:sel}
\int_{2.2}^{5.0} \! PDF(z) \, \mathrm{d}z \geq 0.90,
\end{equation}

\noindent i.e. all the galaxies with a probability of 90\% or higher of being at the redshift range that we are interested in. This leads to a sample of a total of 9203 high redshift galaxies.

\begin{figure}
\centering
\includegraphics[width=0.9\linewidth]{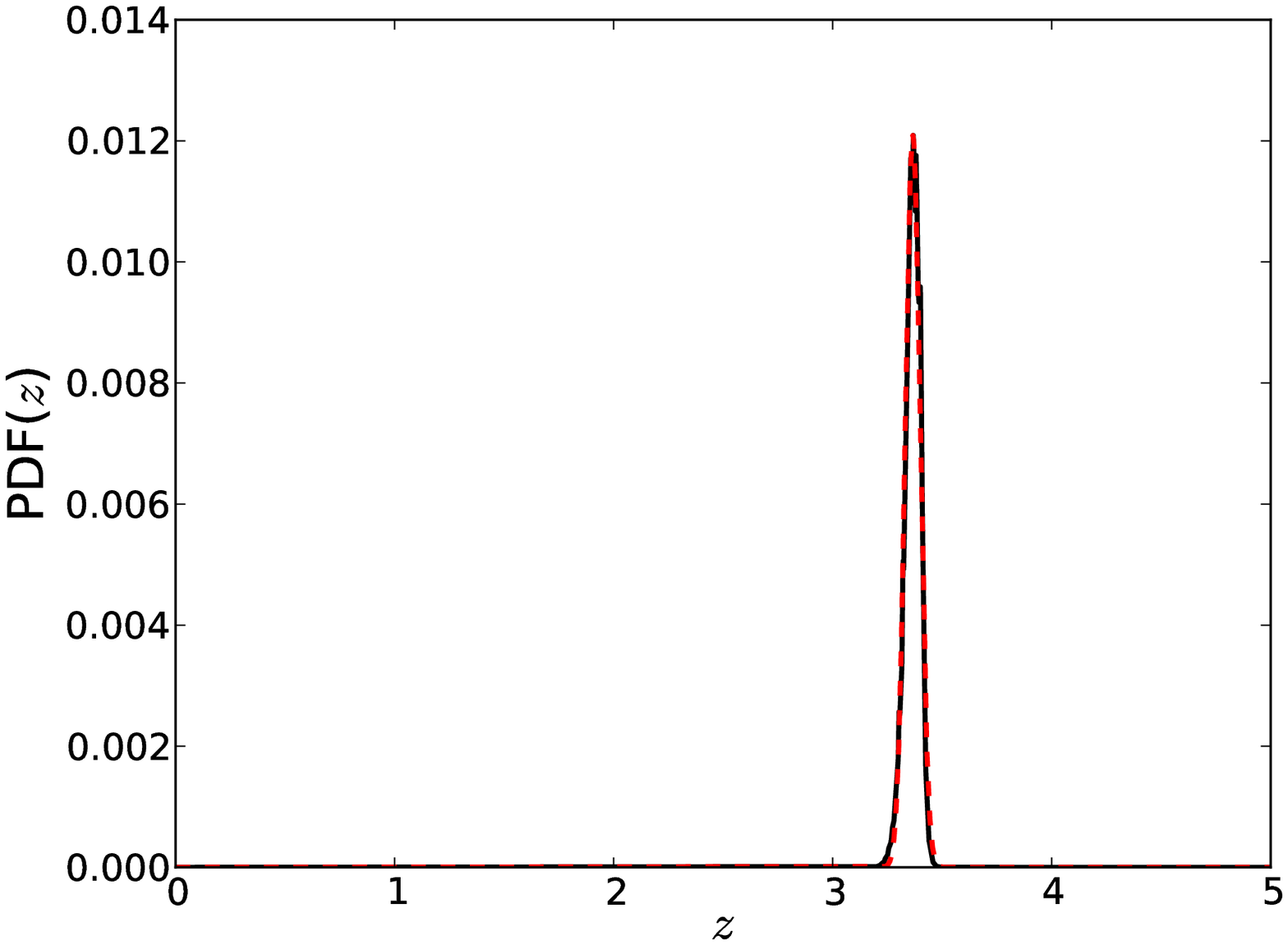}
\includegraphics[width=0.9\linewidth]{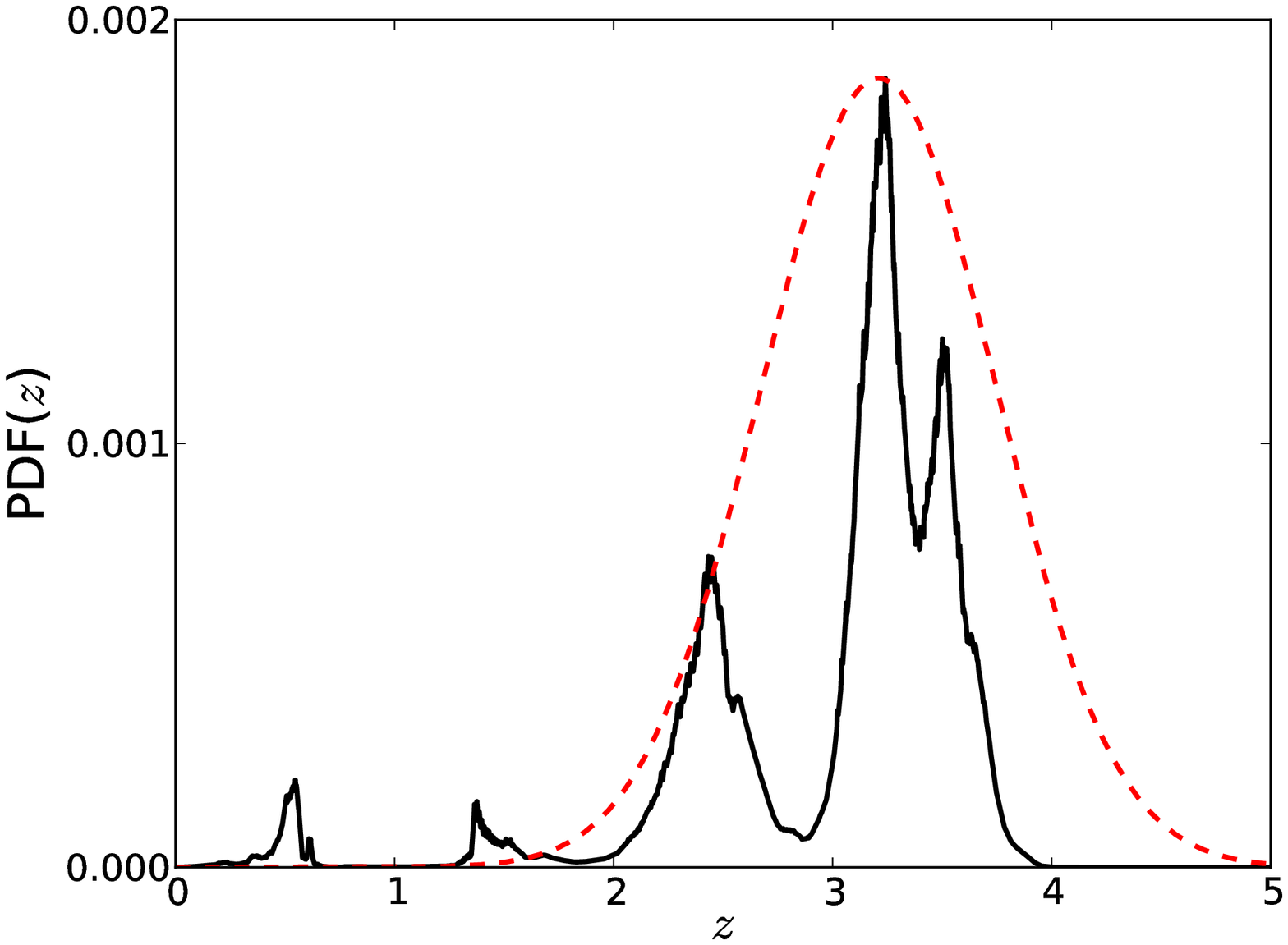}
\caption{zPDFs for two galaxies with very different {\it ML Odds} (black lines), {\bf Top:} {\it ML Odds}=0.898, {\bf Bottom:} {\it ML Odds}=0.084. Overplotted are the corresponding Gaussian approximations of the distributions (dashed red lines). [{\it A colour version of this figure is available in the online edition.}]}
\label{fig:odds}
\end{figure}

\begin{figure}
\centering
\includegraphics[width=0.9\linewidth]{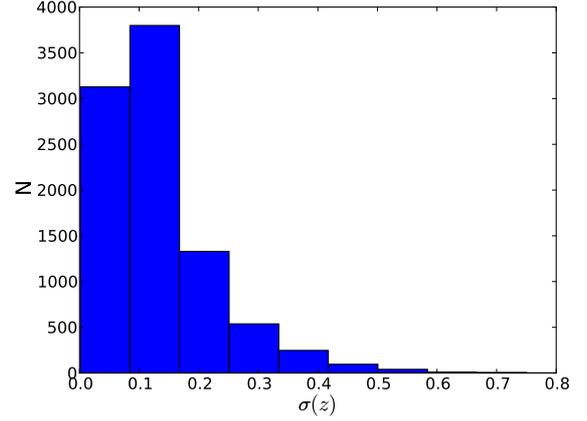}
\caption{The redshift error distribution for a Gaussian approximation for our sample of galaxies at high redshift.}
\label{fig:sigma}
\end{figure}

We note that methodologically this selection could  easily be further refined, if needed. One way would be to study the concentration of the probability distribution around its peak value, e.g. by calculating the ML analogy for the {\it Odds} parameter offered by the BPZ \citep{benitez00}. The {\it Odds} quality parameter is a proxy for the photometric redshift reliability of the sources. The {\it Odds} parameter is defined as the redshift probability enclosed on a $\pm K(1+z)$ region around the main peak in the zPDF of the source, where the constant K is specific for each photometric survey. Molino et al. (2014) find that K=0.0125 is the best value for ALHAMBRA since this is the expected averaged accuracy for most galaxies in the survey. Thus, {\it Odds} $ \in [0,1]$ and it is related to the confidence of the photometric redshifts, making it possible to derive high quality samples with better accuracy and a lower rate of catastrophic outliers. As an example, in Fig.~\ref{fig:odds} we show two zPDFs satisfying  criterion (\ref{eq:sel}), but with very different {\it ML Odds} parameters. We also show  the Gaussian approximations of the corresponding redshift distributions. While it is obvious that a selection in {\it Odds} can refine the redshift selection, we do not make any further selection of this kind. We do not need such a high precision in redshift in order to reject the objects like the one in the bottom panel of Fig.~\ref{fig:odds}. Instead, for all of the galaxies that pass our selection criterion, we calculated the $\sigma$ of their redshift distribution for the Gaussian approximation. The  $\sigma$ distribution for the sample galaxies is shown in Fig.~\ref{fig:sigma}. The average and median of this distribution are 0.13 and 0.11, respectively. We consider this precision to be high enough. Alternatively, in the probabilistic approach (Sect.~\ref{sec:prob}), the whole redshift distribution is taken into account in a natural way; however, we will come back to the question of {\it Odds} in   Sect.~\ref{sec:prob}. We note that in addition to the random errors, we can expect to have systematic errors in the derived redshifts, as was discussed in Sect.~\ref{sec:alhz} and shown in Fig.~\ref{fig:hists} and Table~\ref{tab:z}. However, considering the expected size of these biases, when working with coarse redshift bins, this should not be a  problem.

\subsection{Completeness and contamination}\label{sec:cc}

We estimate here the expected level of contamination and incompleteness of our sample within the limiting magnitude of ALHAMBRA and assuming that the zPDFs correctly reflect the uncertainties in the redshift estimations.

\subsubsection{Contamination}

Presuming the assumption of a flat prior is true, the upper limit for the contamination is directly set by our selection criterion: as we select the objects with a $\geq 90$\% probability of being at desired redshift, we automatically allow a contamination of $\leq10$\% by galaxies at other redshifts. To get a more exact value of the expected contamination, we summed the probabilities of the objects selected by  criterion (\ref{eq:sel}) within the redshift interval $2.2 \leq z \leq 5.0$. The resultant average probability is $\sim$96.5\%, meaning that we could expect a contamination by lower redshift galaxies of only 3.5\%. We can expect a low level of contamination as our selection criterion (\ref{eq:sel}) is rather strict; there may be galaxies with, e.g., a $>50$\% probability of being within our redshift bin but which are not selected by our criterion. This naturally leads to a low level of completeness in our sample as will be discussed in the next section. 

We expect that the most significant source of contamination of our high redshift galaxy sample are the faint red galaxies at low redshifts because of the confusion between the 4000~\AA~and Lyman breaks. In addition, some faint cold stars may be included, as the preselection against stars is statistical and not defined for magnitudes fainter than $m\sim 22.5$ \citep{molino14}, and the noisy spectra of cold stars could be confused with the LBG spectra.

\subsubsection{Completeness}\label{sec:coness}

The completeness at a given redshift bin is defined as the ratio of galaxies at the corresponding redshift that are detected and that also pass the selection criteria to all the rest frame UV-bright galaxies at the given redshift bin actually present in the Universe. It has been shown in \citet{molino14} that the ALHAMBRA catalogues are $\sim 100$\% complete up to $m=24$ in the $F814W$ detection filter. For the high redshift galaxies that we are interested in, this filter traces the UV continuum redwards of the Ly-$\alpha$, the Ly-$\alpha$ break only slightly entering the $F814W$ passband at $z=5$ (see Fig.~\ref{fig:filts}). Hence, considering the UV continuum of the LBGs are generally  flat ($F_{\nu}$=constant), we can also expect  a complete  detection up to $m=24$ in the first filter towards the Ly-$\alpha$ line for the galaxies that we are interested in.

If the flat prior assumption is correct, the expected completeness due to our candidate selection can be derived by summing the probability distributions within the redshift interval that we are interested in for all the objects in our cleaned catalogue which do not fulfil our selection criterion, i.e.

\begin{equation}
N=\sum_i \int_{2.2}^{5.0} \! PDF_i(z) {\rm d}z
\end{equation}

\noindent for all the objects $i$ fulfilling the criterion

\begin{equation}
\int_{2.2}^{5.0} \! PDF_i(z) {\rm d}z<0.9.
\end{equation}

\noindent This sum gives the expected total number of galaxies that are located in the redshift range that we are interested in, but   not selected as such by our criterion. The total number is 40166.8, i.e. $\sim 4.4$ times the objects in our sample. The completeness could be made higher by relaxing the criterion (\ref{eq:sel}), at the cost of increasing the contamination. To carry out statistical studies on the high redshift galaxy population, we certainly should find a better compromise between the contamination and completeness. However, we will discuss later on how statistical studies can be carried out using  the zPDFs directly without any previous candidate selection. Hence, we stick to this candidate selection, which we know to be clean but incomplete, and we name it the clean sample.

\subsubsection{Quasars}\label{sec:qso}

Quasar  spectra are not included in the BPZ spectral templates. Hence, our selection can contain quasars, but we do not expect a complete selection of quasars. We tested if the known quasars observed by ALHAMBRA would fulfil our redshift selection criterion~(\ref{eq:sel}). In total we found 205 ALHAMBRA objects that had counterparts identified as quasars with spectroscopic redshift in other surveys. They consist of 170 
sources from \citet{matute12} (see also references therein), one quasar at $z=5.41$ from \citet{matute13}, 15 sources from the SDSS quasar catalogue DR10 \citep{paris14}, and 19 X-ray sources from CHANDRA that have an associated  optical and infrared counterpart \citep{civano12}. For the CHANDRA sources we also demanded that they were classified as point sources and their variability parameter was greater than 0.25, in agreement with \citet{salvato09}. Of these 205 quasars, 48 have a spectroscopic redshift in the range that we are interested in ($2.2 \leq z \leq 5.0$) and 2 are at higher redshifts ($z=5.07$ and $z=5.41$), while the rest are located at lower redshifts. Of the objects at the redshift interval that we are interested in, 19 (40\%) fulfil our z-selection criteria. In addition, 11 out of the remaining 155 objects at lower-z (7\%) enter our z-selection. The quasar at $z=5.07$ is placed at $z=4.88\pm0.03$, i.e. also enters our redshift selection, and the quasar at $z=5.41$ is placed at $z=5.30\pm0.02$, if Gaussian approximations are used (which in these cases is a good approximation as the redshift distributions show only one significant peak). However, the stellarity flag removes most of the quasars from our final sample so that in the end only five high redshift quasars (10\%) and two (1.3\%) lower redshift quasars enter our sample. We expect to have a better control of these objects once the ALHAMBRA quasar catalogue is available (Chaves-Montero et al., in prep.).

To  get an estimation of the maximum expected contamination of quasars in our clean sample, we compared the $i-$band number counts of quasars at the redshift range $z= 2.2 - 3.5$ \citep{ross13} to the total number of objects in our sample at the same redshift range. The redshift range $z\simeq 2-3$ is often known as the quasar epoch \citep{croom09}, as this is where the number density of bright quasars peaks. Hence, the comparison at this redshift range gives an upper limit of the expected total contamination by high redshift quasars in our sample of high redshift galaxies. From the double power-law fit to the cumulative $i-$band number counts of quasars at $z= 2.2 - 3.5$ \citep{ross13}, a total surface density of 263 deg$^{-1}$ quasars with $m_i <=24$ is derived. Hence, in the ALHAMBRA area we would expect to have $2.38 \, \rm{deg} \times 263 \, \rm{deg}^{-1}$ = 626 quasars brighter than $m_i =24$. The total number of galaxies brighter than $m = 24$ in our clean sample at the same redshift bin is 1707. We roughly compare these numbers without considering a $k$-correction between the $i-$band and our ALHAMBRA bands. If 10\% of the quasars at the ALHAMBRA area enter our selection, as we infer from the spectroscopic sample, the total (maximum) rate of contamination of our clean sample by high-redshift quasars is $0.1 \times 626/1707 = 0.037$, i.e. $< 4\%$.

In addition, we showed above that 1.3\% of quasars at $z<2.2$ can be included in our sample. \citet{ross13} also give the prescription to calculate the quasar surface density at the redshift range $1.0 < z < 2.2$, giving 99.6 deg$^{-1}$ quasars with $m_i <=24$. Doubling this to account for (i.e. overestimate for the much smaller volume)  the quasars at the redshift range $0.0 < z < 1.0$, gives an additional maximum contamination of $2\times 99.6 \, \rm{deg}^{-1} \times 2.38 \, \rm{deg} \times 0.013$ = 6 quasars brighter than $m_i =24$ at $0.0 < z < 2.2$ in the ALHAMBRA area. If this is compared to the total amount of galaxies brighter than $m = 24$ in our clean sample (2296 galaxies), an additional maximum contamination rate of 0.3\% is obtained.

\begin{figure*}
\centering
\includegraphics[width=0.7\linewidth]{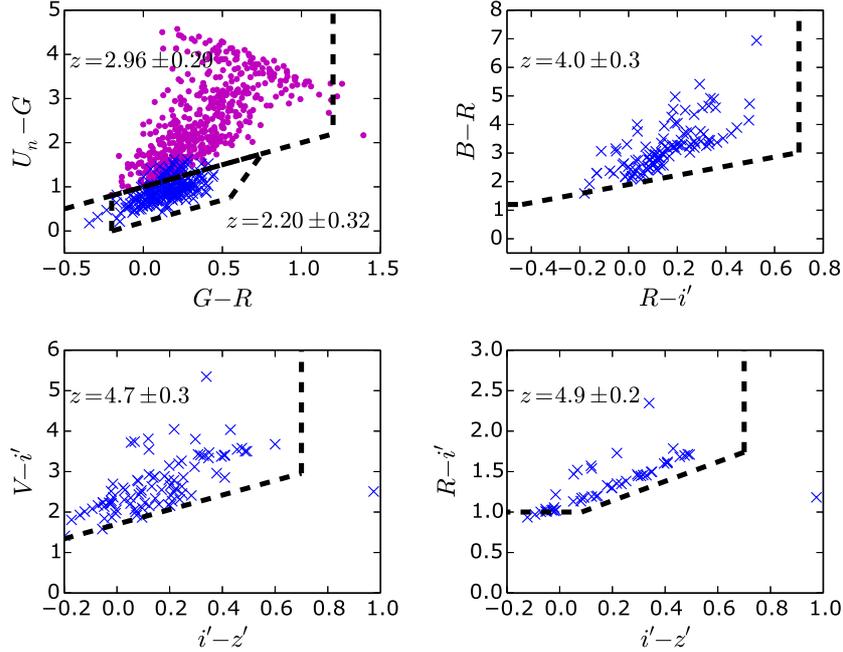}
\caption{Locations of our clean sample candidates in four colour--colour diagrams used for traditional dropout selections. The selection boxes in each diagram are shown with dashed lines and the redshift ranges they target are indicated in each panel. We only plot the candidates in the redshift range within 1$\sigma$ of the one targeted by these diagrams (blue crosses). In the top left diagram the blue crosses refer to the BX selection while the magenta dots to the LBG selection. See the text for more details. [{\it A colour version of this figure is available in the online edition.}]}
\label{fig:colors}
\end{figure*}

\subsection{Comparison with traditional colour selections}\label{sec:comp}

To see if the candidates in our clean sample would have been selected by traditional dropout methods, we tested how they would be located in some traditional colour--colour diagrams. In particular, we opted for testing the BX selection ($\langle z\rangle =2.20\pm0.32$) of \citet{steidel04}; the LBG selection ($\langle z\rangle =2.96\pm0.29$) of \citet{steidel03}; and the {\it BRi$^\prime$} ($\langle z\rangle =4.0\pm0.3$), {\it Vi$^\prime$z$^\prime$} ($\langle z\rangle =4.7\pm0.3$), and {\it Ri$^\prime$z$^\prime$} ($\langle z\rangle =4.9\pm0.2$) LBG selections of \citet{yoshida06}.

First, we carried out SED fitting on our sample galaxies in order to find a spectrum which we  could then convolve with the broadband filters used in the above dropout selections. To assure a good SED-fitting, we considered only the galaxies with good quality photometry in all of the filters by setting "irms\_OPT\_Flag" = 0 and "irms\_NIR\_Flag" = 0. This requirement reduced our sample to 8023 galaxies. For the SED fitting we used the single stellar population (SSP) models of \citet{bruzual03} of all the available metallicities (six metallicity values in the range $Z=0.001-0.05$) and of 40 ages roughly logarithmically spaced from 10 Myr to the age of the Universe. We added  the extinction law of \citet{leitherer02} at the wavelength range 970 \AA-1200 \AA, and that of \citet{calzetti00} for longer wavelengths. At wavelengths below 970 \AA, where neither of the two laws is defined, we adopted a constant extinction with a value equal to that at 970 \AA. The colour excess, $E(B-V)$, was varied in a range of realistic values: from 0.0 to 0.5  \citep{shapley03} in steps of $\Delta E(B-V)=0.025$. The model spectra were moved in redshift in steps of $\Delta z=0.025$ to sample the redshift range that we are interested in, so that at each redshift only the SSPs up to the age of the Universe at that time were considered. The Lyman forest was modelled following the prescriptions of \citet{madau95}, considering the H$\alpha$, H$\beta$, H$\gamma$, and H$\delta$ line blanketing. Below 912 \AA~the opacity was assumed to increase abruptly, leading to practically zero flux bluewards of the Lyman break.

These template spectra were convolved with the ALHAMBRA filter passbands. Each galaxy in our sample was fitted by this template library using the $\chi ^2$-method so that only the templates with redshifts $z_{template}=\langle z \rangle\pm\sigma _z$ were considered, i.e. those templates whose redshift is inside 1$\sigma$ from the median redshift of the fitted galaxy as derived from its zPDF. The template spectrum whose fit produced the lowest value of $\chi ^2$ was then assigned as the best fit template for each galaxy in our sample. Finally, only the galaxies brighter than $m=24$ in the first filter redwards from the Ly-$\alpha$ line and with the reduced $\chi_r ^2 < 2$ ($\chi_r ^2 = \chi ^2/(1-N)$, where $N$ is the number of filters used in the fit) were accepted for the analysis. These steps reduced our sample to 1844 and 1327 galaxies, respectively.

The original spectra of these best fit templates were then convolved with the filter passbands of the broadband filters of interest and the objects were placed in the colour--colour diagrams used in the dropout selections (Fig.~\ref{fig:colors}). To simulate the $G$, $R$, and $U_n$ passband data used in the selections of \citet{steidel03} and \citet{steidel04}, we downloaded the corresponding transmission curves from KPNO website\footnote{https://www.noao.edu/kpno/mosaic/filters/filters.html}. To simulate the selection of \citet{yoshida06}, the $B, R, V, i^\prime$, and $z^\prime$ transmission curves were downloaded from NAOJ website\footnote{http://www.naoj.org/Observing/Instruments/SCam/sensitivity.html}.

In each diagram in Fig.~\ref{fig:colors} we plotted only those candidates of our sample whose (ALHAMBRA median) redshifts are within 1$\sigma$ from the one targeted by the corresponding dropout selections. We see that basically all of our candidates would also be selected  by these traditional colour--colour diagrams. The percentages of the candidates inside the selection boxes are 99\%, 99\%, 97\%, and 94\%, for the LBG, {\it BRi$^\prime$}, {\it Vi$^\prime$z$^\prime$}, and {\it Ri$^\prime$z$^\prime$} selections, respectively. The BX diagram shows the largest scatter outside the selection box, the fraction of candidates inside the box being 83\%. The galaxy clearly outside the selection boxes in the bottom right corners of the bottom diagrams is the same one in both diagrams. It is a very faint object, and even though it is brighter than $m=24$ in the first filter redwards of the Ly-$\alpha$ (the magnitude being $m=23.8$), in all the other filters it is fainter than the 5$\sigma$ limiting magnitude for the corresponding filter.

\begin{figure}
\centering
\includegraphics[width=0.9\linewidth]{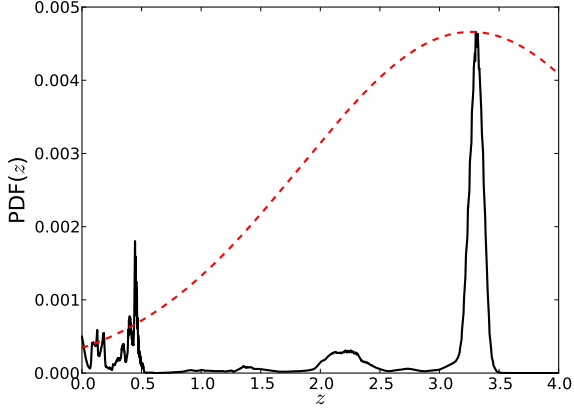}
\caption{Redshift probability distribution of a galaxy with a significant probability at both high and low redshift (black line). Overplotted is the corresponding Gaussian approximation of the distribution (dashed red line). [{\it A colour version of this figure is available in the online edition.}]}
\label{fig:spr}
\end{figure}

\section{Probabilistic approach}\label{sec:prob}

The  selection of the clean sample above is an example of the use of zPDFs when one needs a candidate selection and wants to be certain that the selected galaxies really are at desired redshift. However, selecting both a clean and complete sample is challenging. If one would like to have a more complete sample, one could relax  selection criterion (\ref{eq:sel}). However, relaxing it, for example, to allow all the galaxies with a probability $\geq 50\%$ to be at high redshift to enter the sample would automatically lead  to a contamination rate of $\leq 50\%$ (assuming the flat prior assumption is correct). Hence, for any statistical study one should carefully take care of the incompleteness and contamination corrections.

For many purposes the candidate selection is not needed, but the galaxies and their properties can instead be considered as continua described by their zPDFs. For each catalogued ALHAMBRA object, a zPDF is provided. For some galaxies, as for many objects in our clean sample, this distribution is narrow and could be approximated by a Gaussian distribution without losing much information. However, in other cases the distribution is much more spread out and/or is double peaked. This issue was recently discussed in detail by \citet{lopez-sanjuan14}. In Fig.~\ref{fig:spr} we show an example of a two-peaked and a spread out distribution. Now, if we  claimed the galaxy of Fig.~\ref{fig:spr} to be at any certain redshift bin $z_1 - z_2$ we would certainly fail (unless this bin were wide enough to cover the whole range where the PDF(z) $> 0$). However, for statistical purposes we can interpret the probabilities $p$ of  equation (\ref{eq:pint2}) as fractions. A similar approach was adopted by \citet{mclure09} when deriving LBG luminosity  functions.

\begin{figure*}
\centering
\includegraphics[width=0.7\linewidth]{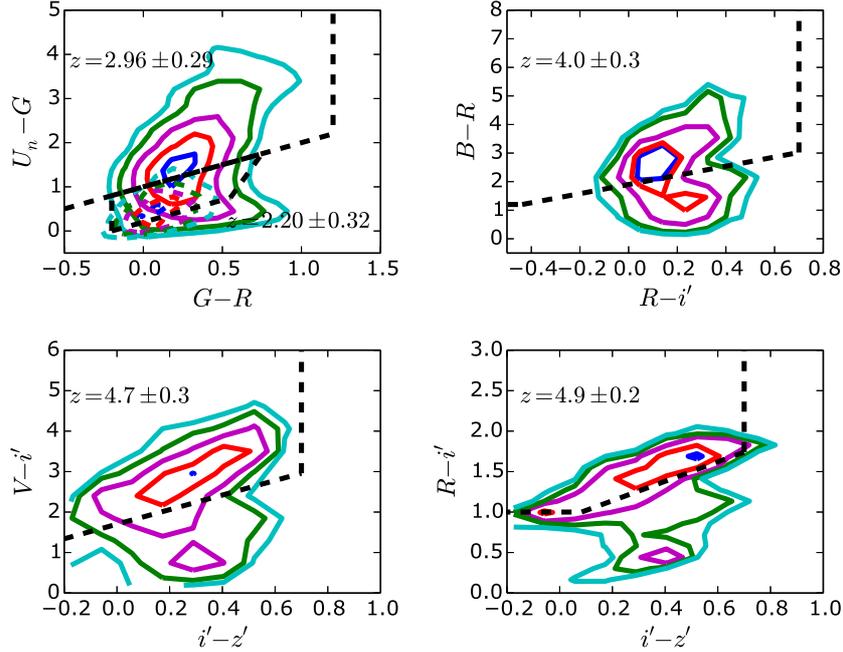}
\caption{Density of the ALHAMBRA high redshift galaxies in four colour--colour diagrams used for traditional dropout selections. The densities are derived using our probabilistic approach. The selection boxes in each diagram are shown with dashed lines, and the redshift ranges they target are indicated in each panel. The contours enclosing 20\%, 40\%, 60\%, 80\%, and 90\% of the objects are marked as solid lines (dashed lines for the BX selection). See the text for more details. [{\it A colour version of this figure is available in the online edition.}]}
\label{fig:colors_all}
\end{figure*}

\subsection{Colour--colour diagrams}\label{sec:ccall}

In Fig.~\ref{fig:colors_all} we show the distribution of the whole set of ALHAMBRA galaxies as density contours in the same colour--colour diagrams as in Fig.~\ref{fig:colors}. To obtain these contours, the catalogue was cleaned as explained in Sect.~\ref{sec:catsel}. In addition, good quality photometry was required in all the filters. All the objects were fitted by \citet{bruzual03} SSP models and convolved with the broadband filter passbands to find their broadband colours as in Sect.~\ref{sec:comp}. Finally, only the galaxies brighter than $m_{\rm{UV}}=24$ in the first filter redwards from the Ly-$\alpha$, and of good quality SED-fitting ($\chi_r < 2$), were used for the analysis (105280 objects). The zPDF of each object was integrated within the redshift interval targeted by each colour--colour diagram (Eq.~\ref{eq:pint2}). To create the density plot in Fig. \ref{fig:colors}, each object was weighted by this fraction. We see that while most of the density of the galaxies lie within the boundaries of the dropout selection boxes, there are also galaxies outside the boxes. The percentages of galaxies outside the boxes are for the BX, LBG, {\it BRi$^\prime$}, {\it Vi$^\prime$z$^\prime$}, and {\it Ri$^\prime$z$^\prime$} selections, respectively, $35\%$, $39\%$, $46\%$, $37\%$, and $39\%$, i.e. more than one third of the restframe UV bright galaxies would be missed by these selections. The existence of galaxies outside the selection boxes supports the known fact that the dropout selections are not complete. A recent spectroscopic study of high redshift galaxies in VUDS survey \citep{lefevre14} also demonstrates the existence of high redshift galaxies outside the $UGR$-selection box, albeit finding a smaller percentage than we did (20\%) of galaxies in the redshift range $2.5 < z < 3.5$ outside the box. On the other hand a similar study in the VVDS survey \citep{lefevre13} reveals that 46\% of the galaxies at the redshift range $2.7 \leq z \leq 3.4$ and with a `reliable' spectral flag are outside the $UGR$-selection box, while 17\% of those with a `very reliable' flag are located outside the box. Of these two surveys, our selection function resembles more that of the VVDS survey (pure magnitude selection) than that of the VUDS where  a photometric redshift selection was also carried out.

\subsection{Number counts}\label{sec:ncounts}

In order to obtain the number $N$ of objects in a redshift bin $z_1 < z < z_2$ and magnitude bin $m_1 < m < m_2$, we carried out a summation over all the objects $i$ in the cleaned ALHAMBRA catalogue of the form

\begin{equation}\label{bnumcounts}
N=\sum_{m_1 < m_i < m_2} \int_{z_{1}}^{z_{2}} \! PDF_i(z) {\rm d}z.
\end{equation}

\begin{figure*}
\centering
\includegraphics[width=0.45\linewidth]{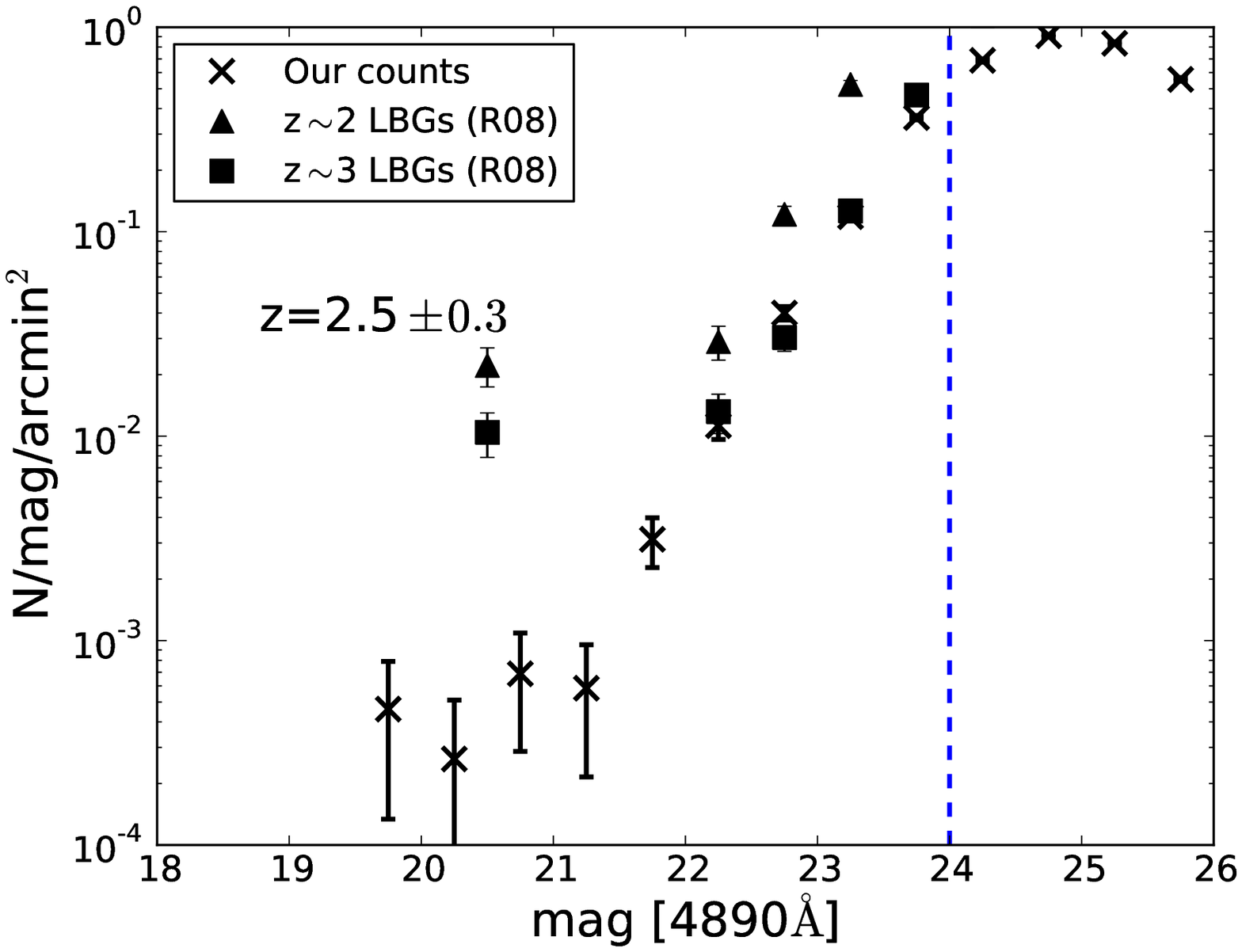}
\includegraphics[width=0.45\linewidth]{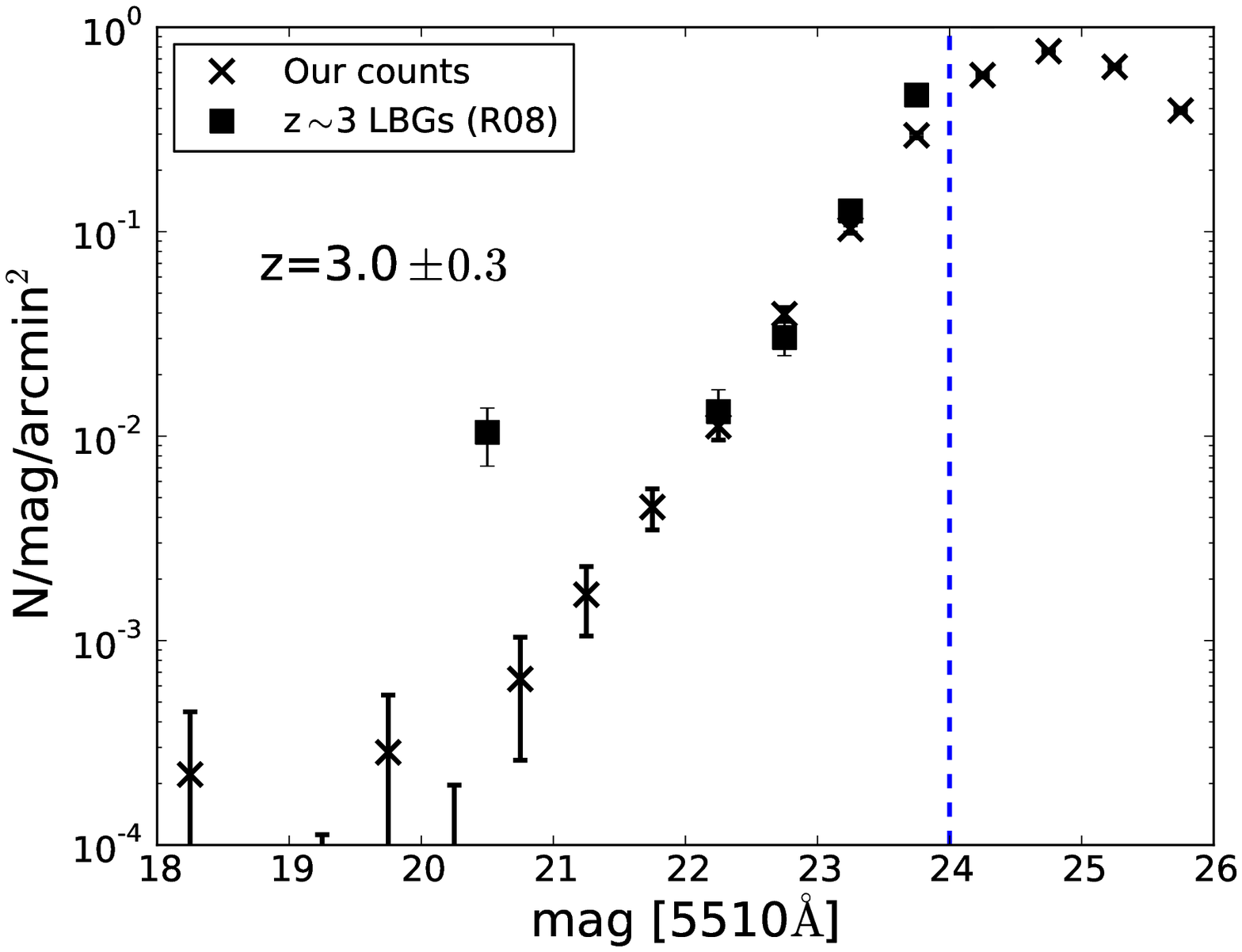}
\includegraphics[width=0.45\linewidth]{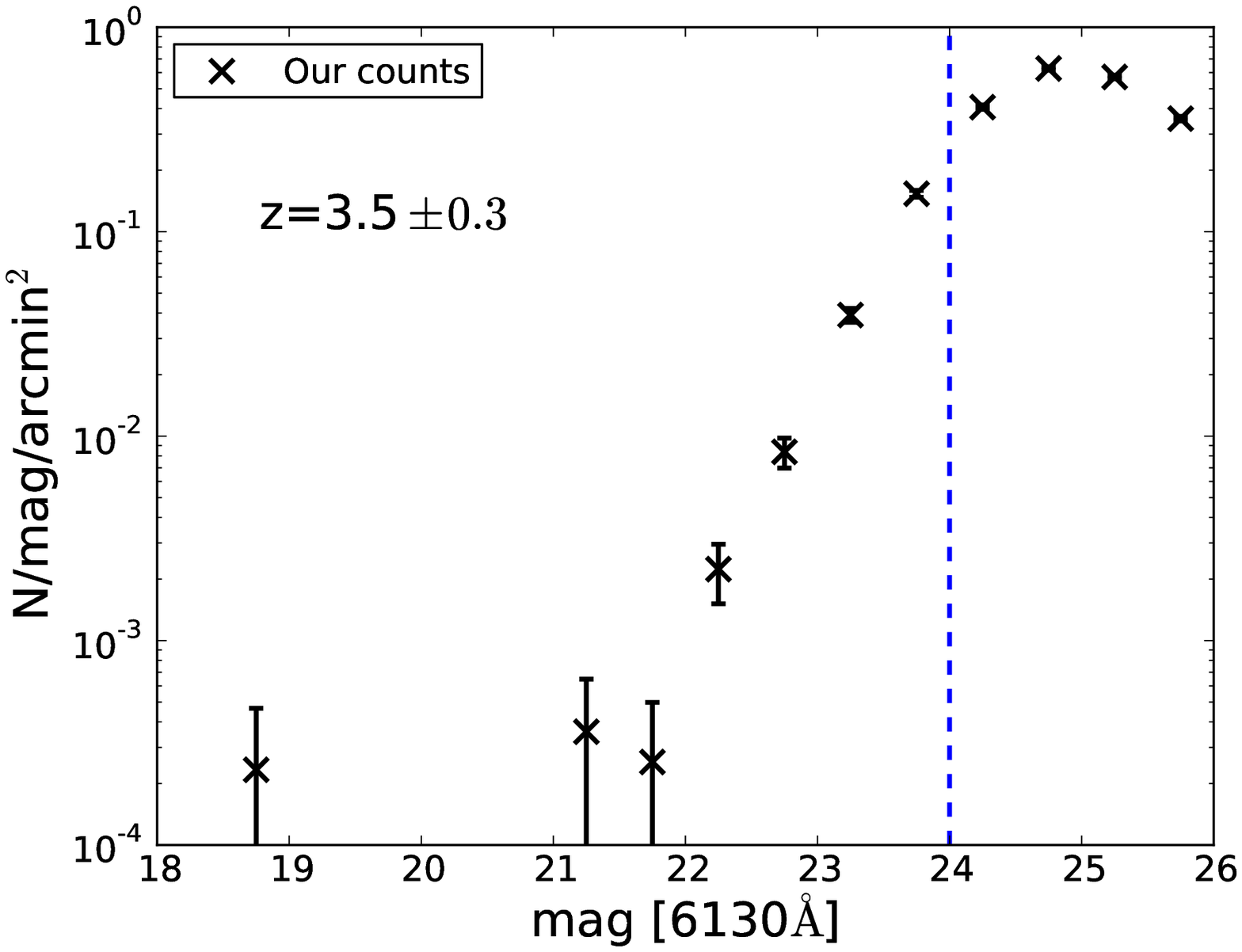}
\includegraphics[width=0.45\linewidth]{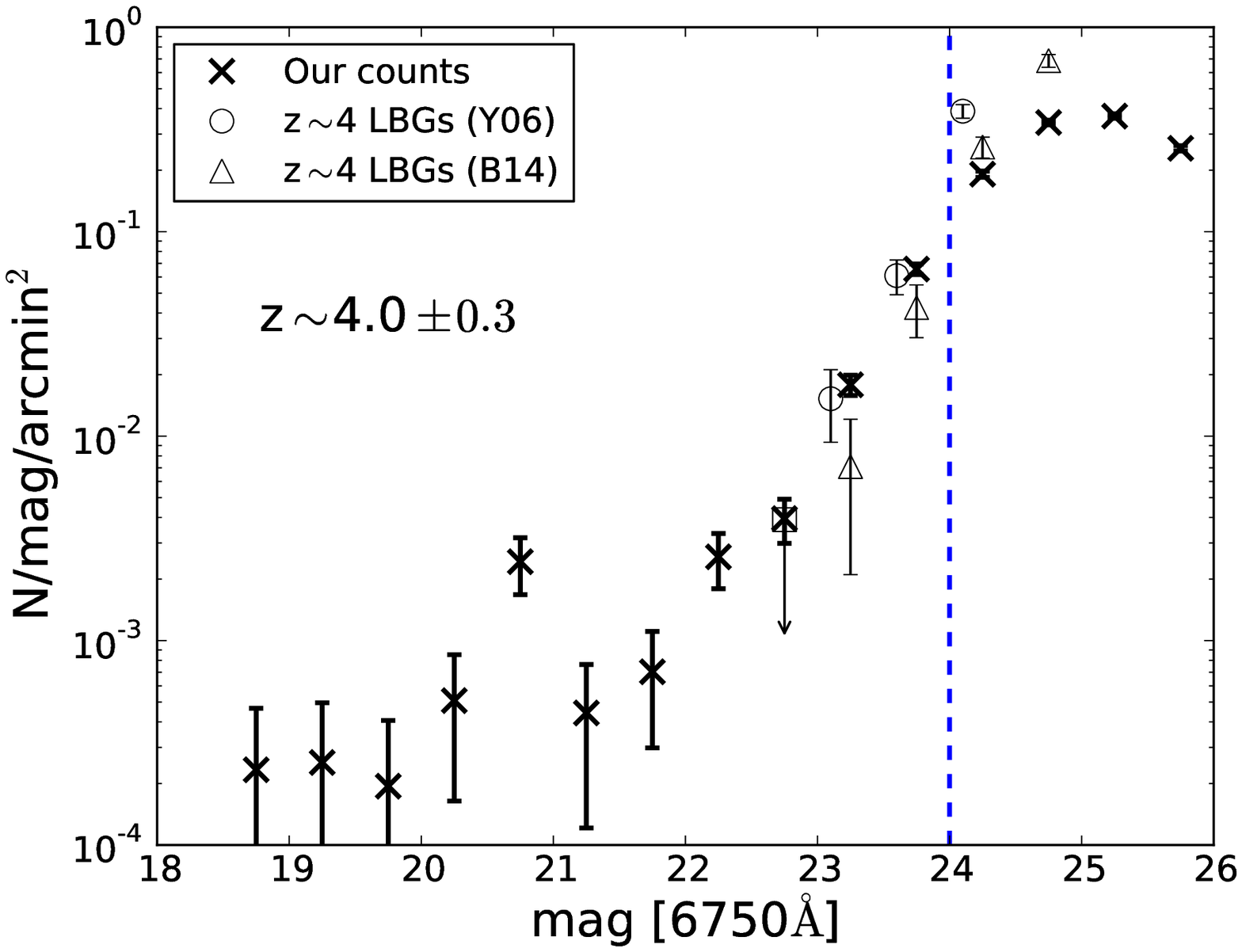}
\includegraphics[width=0.45\linewidth]{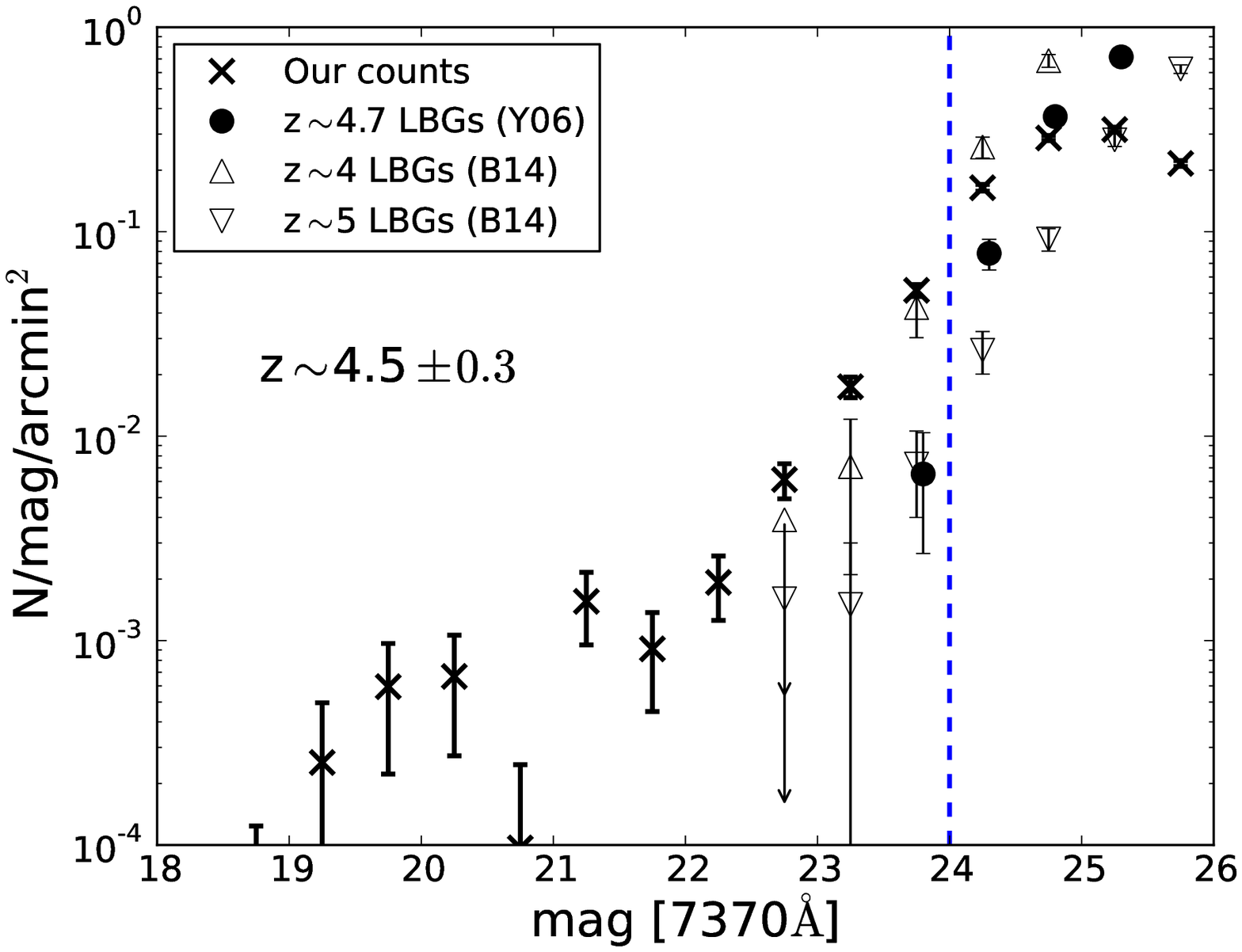}
\caption{Observed number counts for high redshift ALHAMBRA galaxies ({\it crosses}). The error bars reflect Poisson errors. For comparison, we show the BX (z$\sim 2.20\pm0.32$, {\it filled triangles}) and LBG (z$\sim 2.96\pm0.29$, {\it filled squares}) number counts of Reddy et al. (2008; R08), the BRi$^{\prime}$ (z$\sim 4.0\pm0.3$, {\it open circles}) and Vi$^{\prime}$z$^{\prime}$ (z$\sim 4.7\pm0.3$, {\it filled circles}) LBG number counts of Yoshida et al. (2006; Y06), and the $\sim 4$ ({\it open triangles}) and $\sim 5$ ({\it open inverted triangles}) LBG number counts of Bouwens et al. (2014; B14). The ALHAMBRA limiting magnitude is marked at $m=24$ with a blue dashed line. See the text for more details. [{\it A colour version of this figure is available in the online edition.}]} 
\label{fig:ncounts}
\end{figure*}

\noindent For each redshift bin the apparent magnitude refers to the magnitude at the UV continuum as measured by the first filter redwards from the Ly-$\alpha$ (and not containing the possible Ly-$\alpha$ line) at the corresponding redshift. The summation was carried out in five redshift bins. The redshift bins were selected inside the redshift range we consider reliable in our ALHAMBRA data (see Sect.~\ref{sec:alhz}), i.e. $2.2 < z < 5.0$, and we opted for a bin width of $\Delta z=0.6$ to mimic the typical redshift ranges of dropout selected LBGs with which we compare our resulting number counts (see the references in the next paragraph). The resulting probabilistic number counts in bins of 0.5 mag and in redshift bins centred at $z = 2.5, 3.0, 3.5, 4.0,$ and 4.5 in a total area of 8572.5 arcmin$^2$ are shown in Fig.~\ref{fig:ncounts}. These counts are also listed in Table~\ref{tab:pncounts}. 

This method implicitly takes into account both the incompleteness and contamination issues. However,  this method also suffers from quasar contamination as these objects are not considered by the BPZ. We estimated the maximum quasar contamination rate in the same way as in Sect.~\ref{sec:qso} above. We carried out the summation (\ref{bnumcounts}) over all the magnitudes and from $z_1=2.2$ to $z_2= 3.5$ for all non-stellar ALHAMBRA quasars with spectroscopic redshift $z>2.2$ (19 out of 50 quasars). This summation gives 8.27, i.e. $8.27/50=0.165\simeq 17\%$ of the high-redshift quasars contaminate our counts. A similar exercise for  all non-stellar ALHAMBRA quasars with spectroscopic redshift $z<2.2$ (64 out of 155 quasars) leads to 2.48, i.e. 1.6\% of lower redshift quasars contaminating our counts. On the other hand, the total number of ALHAMBRA galaxies brighter than $m = 24$ in the redshift range $2.2 < z < 3.5 $ given by  Eq.~\ref{bnumcounts} is 5269.5. Following the prescription in Sect.~\ref{sec:qso}, this leads to a maximum contamination by high-redshift quasars of $0.165\times 626/5269.5 = 0.019$, i.e. $< 2\%$. The total number of ALHAMBRA galaxies brighter than $m = 24$ in the redshift range $2.2 < z < 5.0 $ given by  Eq.~\ref{bnumcounts} is 5680.9. Hence, the lower-redshift quasars add an additional maximum contamination rate of $2\times 99.6 \, \rm{deg}^{-1} \times 2.38 \, \rm{deg} \times 0.019/5680.9 = 0.0016 < 0.2\%$.

\begin{figure*}
\centering
\includegraphics[width=0.45\linewidth]{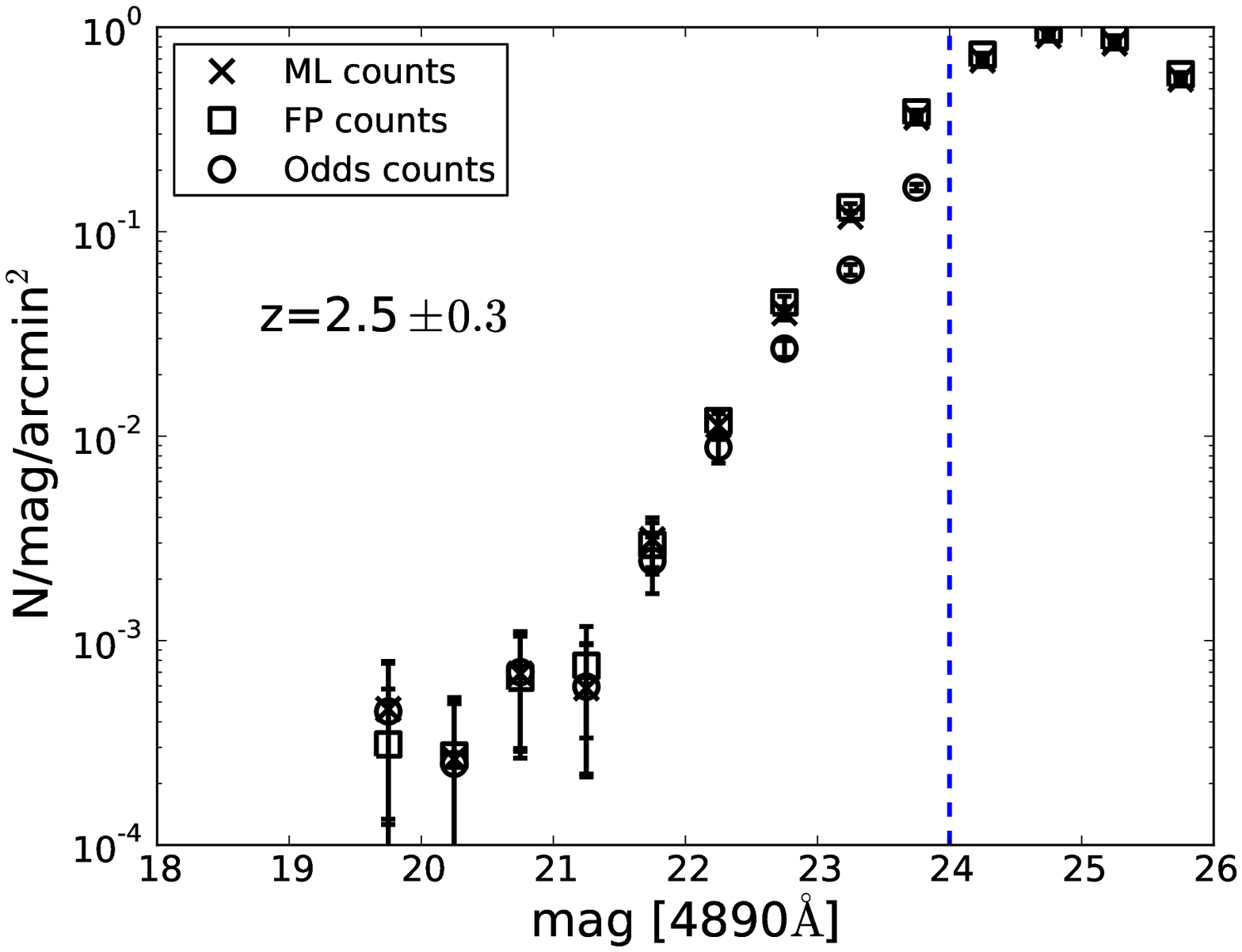}
\includegraphics[width=0.45\linewidth]{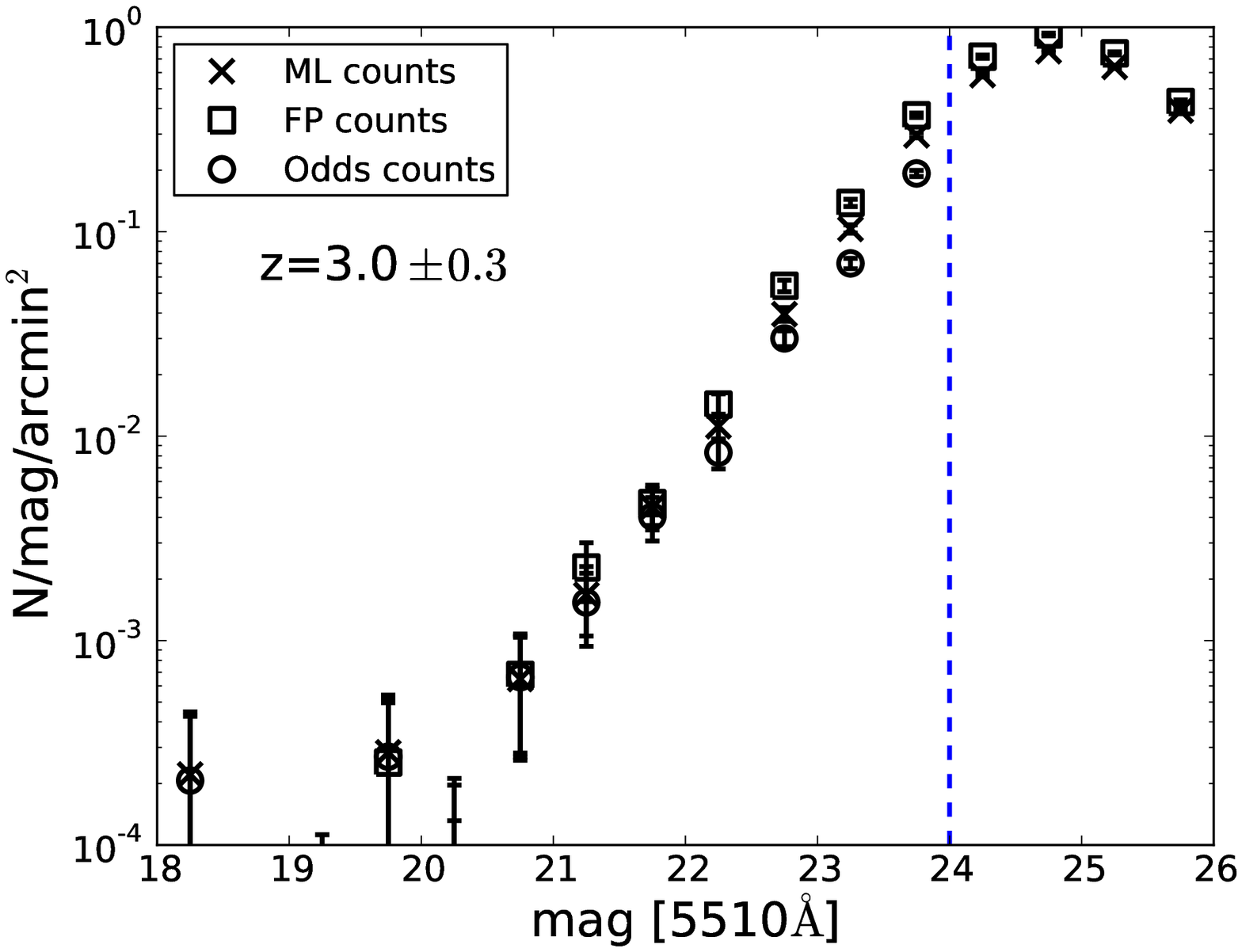}
\includegraphics[width=0.45\linewidth]{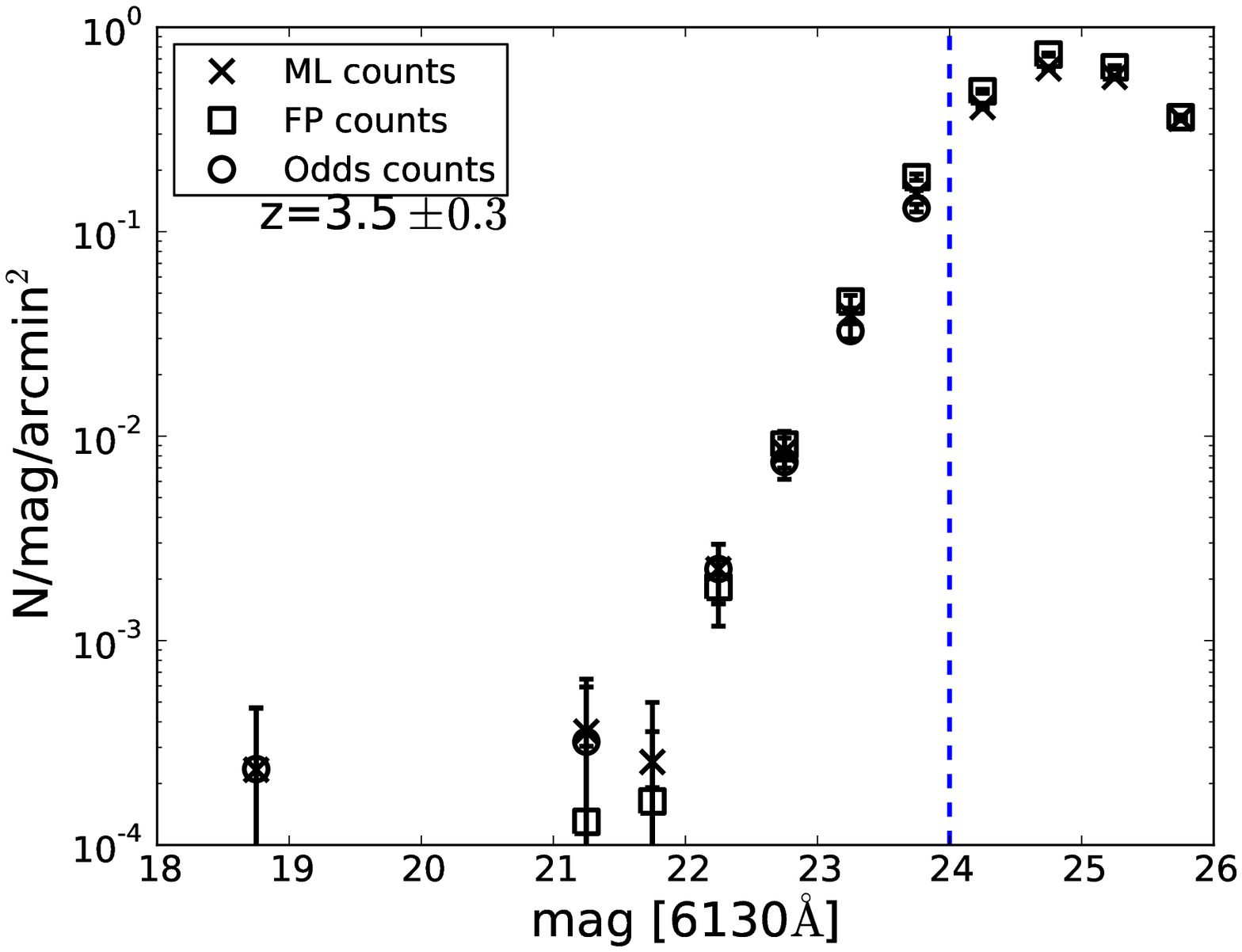}
\includegraphics[width=0.45\linewidth]{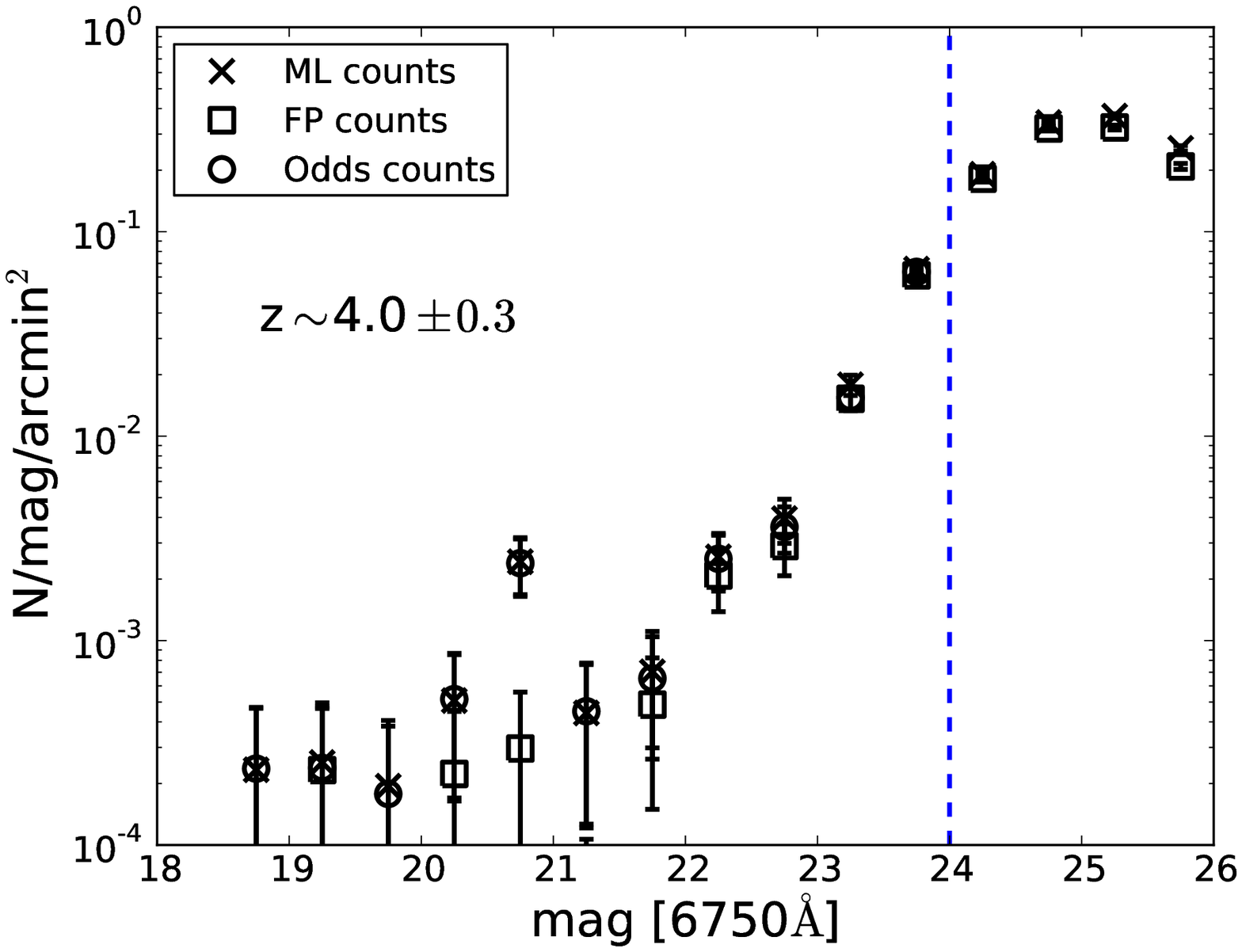}
\includegraphics[width=0.45\linewidth]{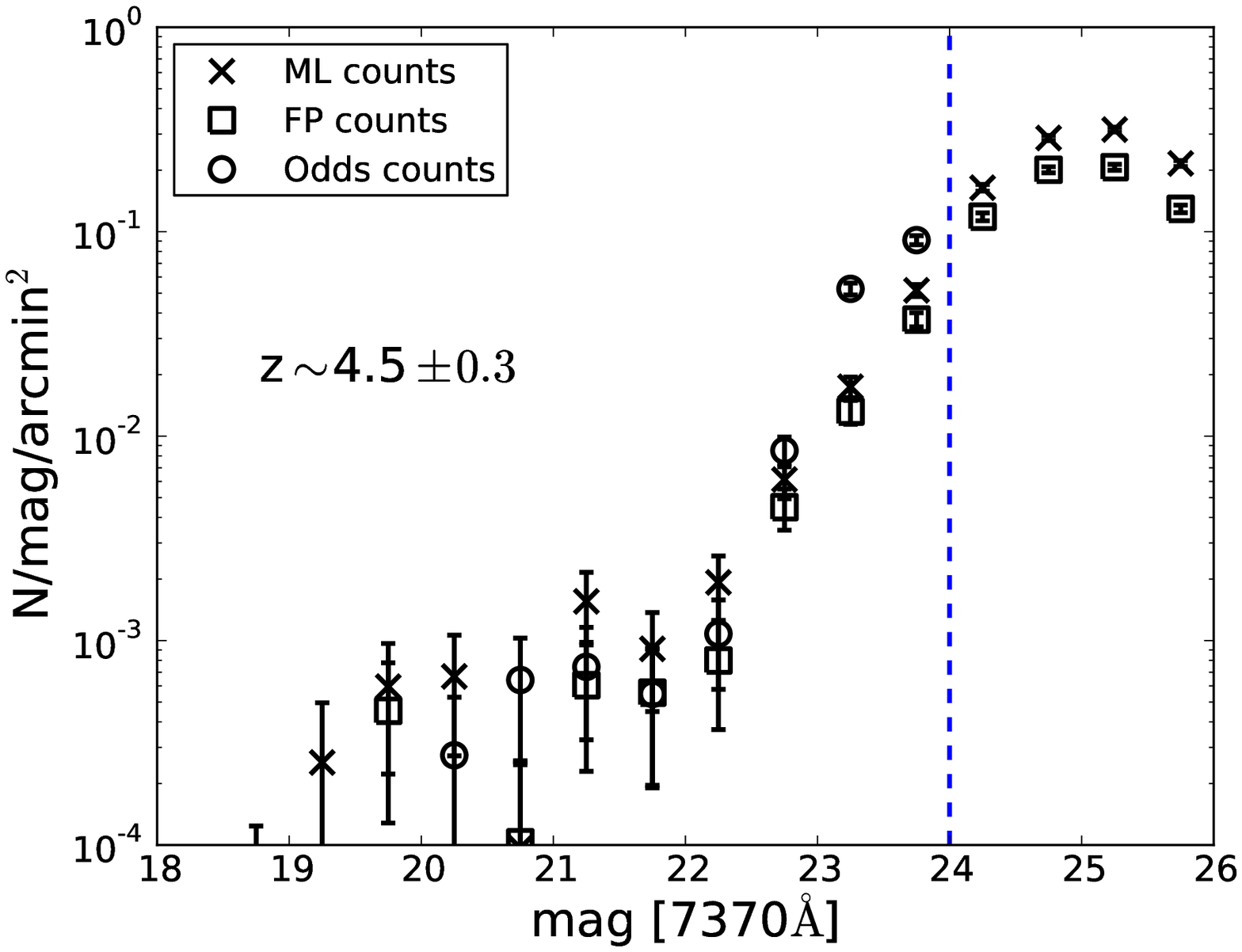}
\caption{Observed probabilistic number counts for high redshift ALHAMBRA galaxies. The ML counts are shown as crosses, the FP counts as open squares, and the counts derived from an "{\it Odds}" selected sample are shown as open circles. The error bars reflect Poisson errors. The ALHAMBRA limiting magnitude is marked at $m=24$ with a blue dashed line. See the text for more details. [{\it A colour version of this figure is available in the online edition.}]} 
\label{fig:ncounts2}
\end{figure*}

\begin{table*}
\caption[]{Probabilistic number counts per magnitude bin at each redshift bin. The total area considered here is 8572.5 arcsec$^2$.}
\label{tab:pncounts}
\centering
\begin{tabular}{c c c c c c c c c c c}
\hline
Magnitude range &N ($z=2.5\pm0.3$) & N ($z=3.0\pm0.3$) & N ($z=3.5\pm0.3$) & N ($z=4.0\pm0.3$) & N ($z=4.5\pm0.3$) \\
\hline
17.0 - 17.5  &   0.0  &    0.0   &    0.1 &    0.1 &   0.0 \\
17.5 - 18.0  &   0.0  &    0.0   &    0.0 &    0.0 &   0.1 \\
18.0 - 18.5  &   0.0 &    0.9   &    0.0 &    0.0 &   0.0 \\
18.5 - 19.0  &   0.0  &    0.0 &    1.0 &    1.0 &   0.1 \\
19.0 - 19.5  &   0.1  &    0.1   &    0.1 &    1.1 &   1.1 \\
19.5 - 20.0  &   2.0  &    1.2   &    0.0 &    0.8 &   2.6 \\
20.0 - 20.5  &   1.1  &    0.3   &    0.1 &    2.2 &   2.9 \\   
20.5 - 21.0  &   2.9  &    2.8   &    0.0 &   10.4 &   0.4 \\ 
21.0 - 21.5  &   2.5  &    7.2   &    1.5 &    1.9 &   6.7 \\ 
21.5 - 22.0  &  13.4  &   19.3   &    1.1 &    3.0 &   3.9 \\ 
22.0 - 22.5  &  48.2  &   47.9   &    9.6 &   11.0 &   8.2 \\ 
22.5 - 23.0  & 171.4  &  169.2   &   35.9 &   16.9 &  26.3 \\
23.0 - 23.5  & 507.8  &  442.5   &  167.6 &   76.4 &  74.7 \\
23.5 - 24.0  & 1549.4 & 1268.7   &  657.2 &  281.2 & 221.4 \\
24.0 - 24.5  & 2950.7 & 2499.5   & 1742.8 &  819.0 &  703.0 \\
24.5 - 25.0  & 3893.4 & 3273.7   & 2694.3 & 1465.7 & 1231.0 \\
25.0 - 25.5  & 3576.8 & 2744.6   & 2449.2 & 1575.9 & 1353.0 \\
25.5 - 26.0  & 2382.2 & 1677.5   & 1531.0 & 1096.2 & 924.4 \\
\hline
\end{tabular}
\end{table*}

For comparison, in Fig.~\ref{fig:ncounts} we have also plotted the dropout selected BX and LBG candidates of \citet[][R08]{reddy08}, and the $BRi^{\prime}$ and $Vi^{\prime}$z$^{\prime}$ dropout selected LBG candidates of \citet[][Y06]{yoshida06}. According to R08, their samples are centred at redshifts $z \sim 2.20\pm0.32$ (BX) and $z \sim 2.96\pm0.29$ (LBG). The LBG samples of Y06 are centred at $z \sim 4.0\pm0.3$ ($BRi^{\prime}$) and $z \sim 4.7\pm0.3$ ($Vi^{\prime}$z$^{\prime}$). We have also overplotted in Fig.~\ref{fig:ncounts} the $z\sim 4$ and $z\sim 5$ dropout selected LBGs of \citet[][B14]{bouwens14}. In our first redshift bin in Fig.~\ref{fig:ncounts} ($z=2.5\pm0.3$) we have plotted both the BX and LBG candidates of \citet{reddy08} as our redshift bin is actually in between the redshift ranges targeted by these two  selections.

\citet{bouwens14} lists the surface densities and their errors in a table (Table 6 in B14) and we have plotted them in Fig.~\ref{fig:ncounts}. The plotted errors for the Y06 and R08 samples reflect the Poisson errors, and we have corrected the Y06 and R08 counts for incompleteness and contamination according to the information given in the corresponding articles: Y06 have studied the completeness and contamination of their sample by simulations. They list the expected number of interlopers for each redshift selection and magnitude bin in tables while the completeness vs. redshift is given in  graphic form for each magnitude bin. For each magnitude bin we opted to adopt the maximum completeness from the distribution for the corresponding magnitude bin. The spectroscopic sample of R08 gives the expected contamination rate for each magnitude and redshift bin, while R08 studied the completeness of their sample (limited to $M_{AB}$ (1700\AA)$< -19.33$) by simulations and found that $\sim 58\%$ of the restframe UV-bright galaxies in the redshift bin $1.9\leq z \leq 2.7$ fulfil the BX colour selection criteria while $\sim 47\%$ of the similar galaxies in the redshift bin $2.7\leq z \leq 3.4$ fulfil the LBG colour selection criteria. In other words, they would expect to find $\sim 42\%$ and $\sim 53\%$ of these galaxies outside the BX and LBG selection boxes, respectively, quite higher fractions than our estimate in  Sect.~\ref{sec:ccall} above.

Detailed comparison of our counts with the counts derived from dropout selections is not straightforward. The dropout selections target a certain redshift range, but a fraction of galaxies from a much wider range of redshift can enter the selections. For example, the BX selection of R08 targets the redshift range $z \sim 2.20\pm0.32$, but the spectroscopic redshift distribution of the galaxies entering the sample, and not considered as contaminants, varies from $z\sim 1.4$ to $z\sim 3.4$. Our methodology simply targets the adopted redshift range. The dropout selections rely on contamination and incompleteness corrections, while our methodology takes these into account implicitly. Despite these differences, the general trends of our counts and the counts from literature coincide. However, in the two lowest redshift bins (centred at $z=2.5$ and $z=3.0$) there is a clear difference between the brightest end of our counts and the brightest bin of R08 counts. The last bin of R08 is wide (from $m=19$ to $m=22$) and we can expect that the counts inside the bin are dominated by the fainter objects. We have plotted their brightest point at the centre of this bin, which slightly exaggerates the difference between our counts and their counts. However, this is not enough to explain the difference. We do not know  where  this difference comes from, but we note that our sampling at the brightest end is clearly better which inclines us to consider our counts more reliable.

Finally, we want to note that in all the redshift bins our counts offer a good sampling of the bright end of the surface densities, down to the magnitudes $m=21-22$. It is also remarkable that according to our counts the total number of ALHAMBRA galaxies brighter than $m_{\rm{UV}}=24$ and at redshifts as high as $\sim4.0$ and $\sim4.5$ is several hundreds, 406 and 348, respectively.

\begin{figure}
\centering
\includegraphics[width=\linewidth]{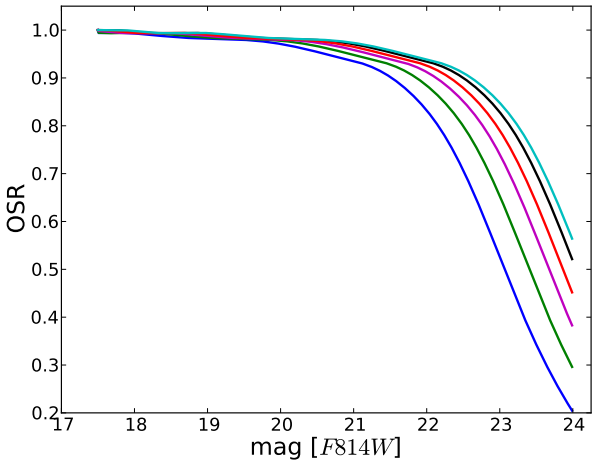}
\caption{The fraction of galaxies with $Odds > 0.3$ as a function of $F814W$ magnitude for different redshift bins: $0.4 < z < 1$ (black line), $z=2.5\pm0.3$ (blue line), $z=3.0\pm0.3$ (green line), $z=3.5\pm0.3$ (magenta line), $z=4.0\pm0.3$ (red line), and $z=4.5\pm0.3$ (cyan line). [{\it A colour version of this figure is available in the online edition.}]} 
\label{fig:osr}
\end{figure}

\subsection{The assumption of a flat prior}

The use of flat (i.e. no prior at all) or very permissive priors in high redshift studies is a common practice \citep[e.g.][]{mclure09,bradley14,lefevre14,duncan14} due to the uncertainties of the prior at high redshift. This means that variation in the density of galaxies as a function of redshift was not considered when deriving the zPDFs. This, in turn, could lead to net contribution of objects from the denser redshift bins to the less dense ones caused by the galaxies with badly defined, flat zPDFs. To test the possible effect of this  on our number counts, we carried out two tests.

First, we derived the counts using a slightly different approach. For each object in the cleaned ALHAMBRA catalogue we calculated its {\it ML Odds} (see Sect.~\ref{sec:sel}) integrating the ML zPDF within the range $z_{\rm ml}\pm 0.0125(1+z_{\rm ml})$, where $z_{\rm ml}$ refers to the redshift of the highest peak of the distribution. Then we eliminated the galaxies with a low {\it ML Odds} value in order to discard the objects with flat zPDFs. To do this, we opted to set {\it ML Odds} $> 0.3$. For these galaxies, we derived their redshift distribution for each magnitude bin and each filter in Fig.~\ref{fig:ncounts} by summing their ML zPDFs. Next, we scaled these redshift distributions to the total number of ALHAMBRA objects in each magnitude bin. From these distributions, we derived the number counts for each redshift bin of interest.

Second,  despite the possible problems the Bayesian prior could cause at the bright end, as was shown in Sect.~\ref{sec:alhz}, we tested how probabilistic number counts would turn out if the FP zPDFs were used. The Bayesian prior takes into account the expected number density variations with redshift and should thus take care of the possible net contributions caused by badly defined, flat zPDFs.

The resulting number counts from the two  experiments described above are plotted in Fig.~\ref{fig:ncounts2} together with the counts derived directly by integrating the ML zPDF of each object. Interestingly indeed, we see that the ML and FP counts roughly coincide. This means that the influence of the prior in the ALHAMBRA high redshift galaxy zPDFs is not significant; i.e. thanks to the ALHAMBRA multifilter system, the ML method alone is capable of recovering  the galaxy redshifts. We do not have enough objects at the very brightest ends of the counts to study if the prior works well there. For this we need to wait for data from larger area surveys.

We see that at the brightest end the results of the Odds experiment coincide with the direct ML counts. At the fainter magnitudes, the {\it Odds} derived counts tend to be lower than the ML (and FP) counts for the two lowest redshift bins (centred at $z=2.5$ and $z=3.0$); in the two following bins (centred at $z=3.5$ and $z=4.0$) all counts coincide in all magnitudes (up to the limiting magnitude), and in the last bin (centred at $z=4.5$) the {\it Odds} derived counts tend to be higher in the faintest magnitudes than the ML/FP counts. This could mean that at the lower redshift bins and fainter magnitudes  a net contribution from low redshift galaxies with flat zPDFs affects our counts and tends to overestimate them, while at the brightest magnitude bins the effect is the opposite. However, considering that the ML and FP counts do agree in these magnitude bins, we do not believe this is the case. The same effect can be obtained if a smaller number of high redshift galaxies have good {\it Odds} values at the first redshift bins than the lower redshift galaxies, and the opposite would be true for the last redshift bin.

To study this in greater  detail, we derived the Odds sampling rate (OSR) as introduced in \citet{lopez-sanjuan14}. This gives the fraction of galaxies with good {\it Odds} values (in this case $Odds > 0.3$) to the total number of galaxies as a function of magnitude in the detection filter, $F814W$. In Fig.~\ref{fig:osr} we show the OSR vs. magnitude derived for $0.4 < z <1$ as a reference curve for lower redshift galaxies, and the corresponding curves for the redshift bins that we are interested in: $z=2.5\pm0.3, 3.0\pm0.3, 3.5\pm0.3, 4.0\pm0.3$, and $4.5\pm0.3$. We see that, at magnitudes fainter than $\sim 20$, the OSR indeed depends on redshift, being lowest for our lowest redshift bin and systematically increasing with redshift, the OSR of our highest redshift bin being higher than that of the reference curve. Actually, this behaviour is also visible in the recovered summed zPDFs of our simulated high redshift galaxies in Sect.~\ref{sec:alhz}. In Fig.~\ref{fig:hists} we see how the recovered summed zPDF becomes narrower with increasing redshift.

To summarise, deriving the galaxy redshift distribution from an {\it Odds} selected sample should be considered with caution as at high redshift OSR strongly depends on redshift. Luckily, we do not need to rely on such an approach as, despite our worries about the use of a prior in our high redshift study (Sect.~\ref{sec:alhz}), the prior does not seem to influence  our counts significantly; both ML and FP zPDFs give similar results. As the prior takes into account the varying galaxy density with redshift, and the FP and ML counts coincide, there clearly is no significant net contribution of objects with spread out zPDFs from the denser redshift bins to the less dense ones. The study of the very brightest and noisy end of our counts (from $m\sim 19$ to $m\sim 21-22$) needs to wait for data from larger area surveys, like J-PLUS and J-PAS. From the number counts derived in this section, we estimate that these surveys will detect tens of thousands of high redshift galaxies brighter than $m=22.5$.

\section{Summary}\label{sec:sum}

So far, most of the studies  of the high redshift UV bright galaxy population have been based on dropout selections. Spectroscopic follow-up of dropout selected samples \citep[e.g.][]{reddy08} have shown that the dropout selection suffers from severe contamination. Simulations \citep[e.g.][]{yoshida06,reddy08} and a spectroscopic study of high redshift galaxies selected from a purely flux-selected sample \citep{lefevre05} have shown that the dropout selection is also highly incomplete. This is further supported by a wide spectroscopic sample by \citet{lefevre14}, where the candidates are selected using photometric redshifts. We expect an alternative probabilistic method, like the one presented here, would help to remove this kind of biases.

We have studied the high redshift UV bright galaxy population in ALHAMBRA data adopting a novel approach based on redshift probability distribution functions (zPDFs). We have shown how a clean sample of high redshift galaxies can be derived from the ALHAMBRA catalogue, integrating the zPDFs and selecting only those galaxies with very high probability to be at high redshift. We studied whether this clean sample would be selected by the traditional dropout techniques, and basically all of the galaxies in our sample actually would also be selected  by these methods at $83 - 99\%$ levels. However, the benefit of our selection  compared to the traditional dropout selections is the expected very low percentage of interlopers.

We have also shown that our clean sample suffers from severe incompleteness and is not able to derive any reliable statistical properties about the high redshift galaxy population. We have introduced a probabilistic method which takes into account both  incompleteness and contamination  in a natural way. In this approach, the galaxies are not treated as unities but rather as fractions in each redshift, where the size of this fraction is derived by integrating the corresponding zPDF of each galaxy at the redshift range of interest. Using this approach, we have studied the distribution of the ALHAMBRA high redshift galaxies in the traditional colour--colour diagrams and discovered that a significant percentage of them ($> 30$\%) are located outside the traditional selection boxes. We have also derived the probabilistic number counts in five redshift bins from $z=2.5$ to $z=4.5$. The strength of our counts is the good sampling of the bright end, down to $m_{\rm{UV}}$(AB)=21-22.

In our simulation we discovered that if the Bayesian prior was used, very bright high redshift galaxies would systematically not be selected to form part of our clean sample. For this reason, all the above studies are based on maximum likelihood (ML) zPDFs, i.e. throughout the paper we have assumed a flat prior. However, we tested how the probabilistic number counts would turn out if the full probability (FP) zPDFs were used. We found  that the FP and ML counts closely match. This reinforces the reliability of these counts. To know if this holds at the very brightest magnitudes, where our data is dominated by noise, data from still larger area surveys is needed. We would like to be able to come back to this issue once the data from wide area ($\sim 8500$ deg$^2$) J-PLUS and J-PAS multifilter surveys are available. From the number counts derived in this work, we estimate that we could detect tens of thousands of high redshift galaxies brighter than $m_{\rm{UV}}$(AB)=22.5 in these surveys. 

We also repeated the probabilistic number counts calculation deriving the ML redshift distributions in each magnitude bin from a selection of galaxies with well-defined photometric redshifts ({\it ML Odds} $> 0.3$) and scaling these to the total number of objects in each magnitude bin. In the faintest magnitude bins we find differences between these and the direct ML counts. Considering that the FP and ML counts roughly coincide in all magnitude bins, we inferred that the differences seen with the counts derived from the {\it Odds} selected sample are due to variations in the fractional amount of galaxies with good {\it Odds} values with redshift. We studied the evolution of this fraction as a function of magnitude and redshift, and found out that at faint magnitudes this fraction indeed varies with redshift.

Even though we have discussed here only the application of the probabilistic method of deriving the galaxy number counts, a similar approach could be used to study any redshift dependent galaxy property. \citet{mclure09} used a similar approach to derive LBG luminosity functions and, recently, Lopez-Sanjuan et al. (2014) discussed a similar approach to study the galaxy merger fraction. We will further study the ALHAMBRA high redshift galaxies using this methodology in the forthcoming papers.

Theoretically, our probabilistic method is totally free of biases due to incompleteness and contamination. However,  this is not totally true, as our photo-z estimations are limited to what is already known about the galaxy population because empirical templates are used. The way to improve  this aspect of the method is to create unbiased lists of candidates, spectroscopically confirm them, and consequently refine the high redshift templates. For the unbiased candidate selection, zPDFs offer a unique opportunity. A wide spectroscopic campaign on candidates selected using zPDFs is already in progress \citep{lefevre14}. Once the spectra of objects derived from unbiased samples are available, these can  be used to improve the photo-z estimations at high redshift, and subsequently improve the accuracy of the statistical methods like the one presented here.

\begin{acknowledgements}
We acknowledge the anonymous referee for the useful comments. K. Viironen acknowledges the Juan de la Cierva fellowship of the Spanish government. We acknowledge funding from the FITE (Fondos de Inversiones de Teruel) and support from the Spanish Ministry for Economy and Competitiveness and FEDER funds through grants AYA2012-30789, AYA2006-14056, AYA 2003-00128, AYA 2006-01325, AYA 2007-62190, AYA2010-15169, AYA2010-22111-C03-02 and AYA2013-48623-C2-2. We also acknowledge Junta de Andaluc\'ia through the grant TIC 114 and Generalitat Valenciana projects Prometeo 2009/064 and PROMETEOII/2014/060, and the financial support from the Arag\'on Government through the Research Group E103. I. Oteo acknowledges support from the European Research Council (ERC) in the form of Advanced Grant, {\sc cosmicism}. A. J. Cenarro  acknowledges the Ram\'on y Cajal fellowship of the Spanish government. M. Povi\'c acknowledges financial support from JAE-Doc program of the Spanish National Research Council (CSIC), co-funded by the European Social Fund. This research made use of Matplotlib, a 2D graphics package used for Python for publication-quality image generation across user interfaces and operating systems (Hunter 2007).    
\end{acknowledgements}

\bibliographystyle{aa}
\bibliography{refs}

\begin{thebibliography}{54}
\expandafter\ifx\csname natexlab\endcsname\relax\def\natexlab#1{#1}\fi

\bibitem[{{Aparicio Villegas} {et~al.}(2010){Aparicio Villegas}, {Alfaro},
  {Cabrera-Ca{\~n}o}, {Moles}, {Ben{\'{\i}}tez}, {Perea}, {del Olmo},
  {Fern{\'a}ndez-Soto}, {Crist{\'o}bal-Hornillos}, {Husillos}, {Aguerri},
  {Broadhurst}, {Castander}, {Cepa}, {Cervi{\~n}o}, {Gonz{\'a}lez Delgado},
  {Infante}, {M{\'a}rquez}, {Masegosa}, {Mart{\'{\i}}nez}, {Prada}, {Quintana},
  \& {S{\'a}nchez}}]{aparicio-villegas12}
{Aparicio Villegas}, T., {Alfaro}, E.~J., {Cabrera-Ca{\~n}o}, J., {et~al.}
  2010, \aj, 139, 1242

\bibitem[{{Arnalte-Mur} {et~al.}(2014){Arnalte-Mur}, {Mart{\'{\i}}nez},
  {Norberg}, {Fern{\'a}ndez-Soto}, {Ascaso}, {Merson}, {Aguerri}, {Castander},
  {Hurtado-Gil}, {L{\'o}pez-Sanjuan}, {Molino}, {Montero-Dorta}, {Stefanon},
  {Alfaro}, {Aparicio-Villegas}, {Ben{\'{\i}}tez}, {Broadhurst},
  {Cabrera-Ca{\~n}o}, {Cepa}, {Cervi{\~n}o}, {Crist{\'o}bal-Hornillos}, {del
  Olmo}, {Gonz{\'a}lez Delgado}, {Husillos}, {Infante}, {M{\'a}rquez},
  {Masegosa}, {Moles}, {Perea}, {Povi{\'c}}, {Prada}, \&
  {Quintana}}]{arnalte14}
{Arnalte-Mur}, P., {Mart{\'{\i}}nez}, V.~J., {Norberg}, P., {et~al.} 2014,
  \mnras, 441, 1783

\bibitem[{{Barger} {et~al.}(2008){Barger}, {Cowie}, \& {Wang}}]{barger08}
{Barger}, A.~J., {Cowie}, L.~L., \& {Wang}, W.-H. 2008, \apj, 689, 687

\bibitem[{{Ben{\'{\i}}tez}(2000)}]{benitez00}
{Ben{\'{\i}}tez}, N. 2000, \apj, 536, 571

\bibitem[{{Benitez} {et~al.}(2014){Benitez}, {Dupke}, {Moles}, {Sodre},
  {Cenarro}, {Marin-Franch}, {Taylor}, {Cristobal}, {Fernandez-Soto}, {Mendes
  de Oliveira}, {Cepa-Nogue}, {Abramo}, {Alcaniz}, {Overzier},
  {Hernandez-Monteagudo}, {Alfaro}, {Kanaan}, {Carvano}, {Reis}, {Martinez
  Gonzalez}, {Ascaso}, {Ballesteros}, {Xavier}, {Varela}, {Ederoclite},
  {Vazquez Ramio}, {Broadhurst}, {Cypriano}, {Angulo}, {Diego}, {Zandivarez},
  {Diaz}, {Melchior}, {Umetsu}, {Spinelli}, {Zitrin}, {Coe}, {Yepes}, {Vielva},
  {Sahni}, {Marcos-Caballero}, {Shu Kitaura}, {Maroto}, {Masip}, {Tsujikawa},
  {Carneiro}, {Gonzalez Nuevo}, {Carvalho}, {Reboucas}, {Carvalho}, {Abdalla},
  {Bernui}, {Pigozzo}, {Ferreira}, {Chandrachani Devi}, {Bengaly}, {Campista},
  {Amorim}, {Asari}, {Bongiovanni}, {Bonoli}, {Bruzual}, {Cardiel}, {Cava},
  {Cid Fernandes}, {Coelho}, {Cortesi}, {Delgado}, {Diaz Garcia}, {Espinosa},
  {Galliano}, {Gonzalez-Serrano}, {Falcon-Barroso}, {Fritz}, {Fernandes},
  {Gorgas}, {Hoyos}, {Jimenez-Teja}, {Lopez-Aguerri}, {Lopez-San Juan},
  {Mateus}, {Molino}, {Novais}, {OMill}, {Oteo}, {Perez-Gonzalez}, {Poggianti},
  {Proctor}, {Ricciardelli}, {Sanchez-Blazquez}, {Storchi-Bergmann}, {Telles},
  {Schoennell}, {Trujillo}, {Vazdekis}, {Viironen}, {Daflon},
  {Aparicio-Villegas}, {Rocha}, {Ribeiro}, {Borges}, {Martins}, {Marcolino},
  {Martinez-Delgado}, {Perez-Torres}, {Siffert}, {Calvao}, {Sako}, {Kessler},
  {Alvarez-Candal}, {De Pra}, {Roig}, {Lazzaro}, {Gorosabel}, {Lopes de
  Oliveira}, {Lima-Neto}, {Irwin}, {Liu}, {Alvarez}, {Balmes}, {Chueca},
  {Costa-Duarte}, {da Costa}, {Dantas}, {Diaz}, {Fabregat}, {Ferrari},
  {Gavela}, {Gracia}, {Gruel}, {Gutierrez}, {Guzman}, {Hernandez-Fernandez},
  {Herranz}, {Hurtado-Gil}, {Jablonsky}, {Laporte}, {Le Tiran}, {Licandro},
  {Lima}, {Martin}, {Martinez}, {Montero}, {Penteado}, {Pereira}, {Peris},
  {Quilis}, {Sanchez-Portal}, {Soja}, {Solano}, {Torra}, \&
  {Valdivielso}}]{benitez14b}
{Benitez}, N., {Dupke}, R., {Moles}, M., {et~al.} 2014, ArXiv e-prints

\bibitem[{{Ben{\'{\i}}tez} {et~al.}(2009){Ben{\'{\i}}tez}, {Moles}, {Aguerri},
  {Alfaro}, {Broadhurst}, {Cabrera-Ca{\~n}o}, {Castander}, {Cepa},
  {Cervi{\~n}o}, {Crist{\'o}bal-Hornillos}, {Fern{\'a}ndez-Soto}, {Gonz{\'a}lez
  Delgado}, {Infante}, {M{\'a}rquez}, {Mart{\'{\i}}nez}, {Masegosa}, {Del
  Olmo}, {Perea}, {Prada}, {Quintana}, \& {S{\'a}nchez}}]{benitez09}
{Ben{\'{\i}}tez}, N., {Moles}, M., {Aguerri}, J.~A.~L., {et~al.} 2009, \apjl,
  692, L5

\bibitem[{{Bouwens} {et~al.}(2014){Bouwens}, {Illingworth}, {Oesch}, {Trenti},
  {Labbe'}, {Bradley}, {Carollo}, {van Dokkum}, {Gonzalez}, {Holwerda},
  {Franx}, {Spitler}, {Smit}, \& {Magee}}]{bouwens14}
{Bouwens}, R.~J., {Illingworth}, G.~D., {Oesch}, P.~A., {et~al.} 2014, ArXiv
  e-prints

\bibitem[{{Bowler} {et~al.}(2014){Bowler}, {Dunlop}, {McLure}, {McCracken},
  {Furusawa}, {Taniguchi}, {Fynbo}, {Milvang-Jensen}, \& {Le Fevre}}]{bowler14}
{Bowler}, R.~A.~A., {Dunlop}, J.~S., {McLure}, R.~J., {et~al.} 2014, ArXiv
  e-prints

\bibitem[{{Bradley} {et~al.}(2014){Bradley}, {Zitrin}, {Coe}, {Bouwens},
  {Postman}, {Balestra}, {Grillo}, {Monna}, {Rosati}, {Seitz}, {Host}, {Lemze},
  {Moustakas}, {Moustakas}, {Shu}, {Zheng}, {Broadhurst}, {Carrasco}, {Jouvel},
  {Koekemoer}, {Medezinski}, {Meneghetti}, {Nonino}, {Smit}, {Umetsu},
  {Bartelmann}, {Ben{\'{\i}}tez}, {Donahue}, {Ford}, {Infante}, {Jimenez-Teja},
  {Kelson}, {Lahav}, {Maoz}, {Melchior}, {Merten}, \& {Molino}}]{bradley14}
{Bradley}, L.~D., {Zitrin}, A., {Coe}, D., {et~al.} 2014, \apj, 792, 76

\bibitem[{{Bruzual} \& {Charlot}(2003)}]{bruzual03}
{Bruzual}, G. \& {Charlot}, S. 2003, \mnras, 344, 1000

\bibitem[{{Calzetti} {et~al.}(2000){Calzetti}, {Armus}, {Bohlin}, {Kinney},
  {Koornneef}, \& {Storchi-Bergmann}}]{calzetti00}
{Calzetti}, D., {Armus}, L., {Bohlin}, R.~C., {et~al.} 2000, \apj, 533, 682

\bibitem[{{Civano} {et~al.}(2012){Civano}, {Elvis}, {Brusa}, {Comastri},
  {Salvato}, {Zamorani}, {Aldcroft}, {Bongiorno}, {Capak}, {Cappelluti},
  {Cisternas}, {Fiore}, {Fruscione}, {Hao}, {Kartaltepe}, {Koekemoer}, {Gilli},
  {Impey}, {Lanzuisi}, {Lusso}, {Mainieri}, {Miyaji}, {Lilly}, {Masters},
  {Puccetti}, {Schawinski}, {Scoville}, {Silverman}, {Trump}, {Urry},
  {Vignali}, \& {Wright}}]{civano12}
{Civano}, F., {Elvis}, M., {Brusa}, M., {et~al.} 2012, \apjs, 201, 30

\bibitem[{{Coe} {et~al.}(2006){Coe}, {Ben{\'{\i}}tez}, {S{\'a}nchez}, {Jee},
  {Bouwens}, \& {Ford}}]{coe06}
{Coe}, D., {Ben{\'{\i}}tez}, N., {S{\'a}nchez}, S.~F., {et~al.} 2006, \aj, 132,
  926

\bibitem[{{Crist{\'o}bal-Hornillos} {et~al.}(2009){Crist{\'o}bal-Hornillos},
  {Aguerri}, {Moles}, {Perea}, {Castander}, {Broadhurst}, {Alfaro},
  {Ben{\'{\i}}tez}, {Cabrera-Ca{\~n}o}, {Cepa}, {Cervi{\~n}o},
  {Fern{\'a}ndez-Soto}, {Gonz{\'a}lez Delgado}, {Husillos}, {Infante},
  {M{\'a}rquez}, {Mart{\'{\i}}nez}, {Masegosa}, {del Olmo}, {Prada},
  {Quintana}, \& {S{\'a}nchez}}]{cristobal-hornillos09}
{Crist{\'o}bal-Hornillos}, D., {Aguerri}, J.~A.~L., {Moles}, M., {et~al.} 2009,
  \apj, 696, 1554

\bibitem[{{Croom} {et~al.}(2009){Croom}, {Richards}, {Shanks}, {Boyle},
  {Strauss}, {Myers}, {Nichol}, {Pimbblet}, {Ross}, {Schneider}, {Sharp}, \&
  {Wake}}]{croom09}
{Croom}, S.~M., {Richards}, G.~T., {Shanks}, T., {et~al.} 2009, \mnras, 399,
  1755

\bibitem[{{Duncan} {et~al.}(2014){Duncan}, {Conselice}, {Mortlock}, {Hartley},
  {Guo}, {Ferguson}, {Dav{\'e}}, {Lu}, {Ownsworth}, {Ashby}, {Dekel},
  {Dickinson}, {Faber}, {Giavalisco}, {Grogin}, {Kocevski}, {Koekemoer},
  {Somerville}, \& {White}}]{duncan14}
{Duncan}, K., {Conselice}, C.~J., {Mortlock}, A., {et~al.} 2014, \mnras, 444,
  2960

\bibitem[{{Finkelstein} {et~al.}(2014){Finkelstein}, {Ryan}, {Papovich},
  {Dickinson}, {Song}, {Somerville}, {Ferguson}, {Salmon}, {Giavalisco},
  {Koekemoer}, {Ashby}, {Behroozi}, {Castellano}, {Dunlop}, {Faber}, {Fazio},
  {Fontana}, {Grogin}, {Hathi}, {Jaacks}, {Kocevski}, {Livermore}, {McLure},
  {Merlin}, {Mobasher}, {Newman}, {Rafelski}, {Tilvi}, \&
  {Willner}}]{finkelstein14}
{Finkelstein}, S.~L., {Ryan}, Jr., R.~E., {Papovich}, C., {et~al.} 2014, ArXiv
  e-prints

\bibitem[{{Guhathakurta} {et~al.}(1990){Guhathakurta}, {Tyson}, \&
  {Majewski}}]{guhathakurta90}
{Guhathakurta}, P., {Tyson}, J.~A., \& {Majewski}, S.~R. 1990, \apjl, 357, L9

\bibitem[{{Le Fevre} {et~al.}(2013){Le Fevre}, {Cassata}, {Cucciati}, {de la
  Torre}, {Garilli}, {Ilbert}, {Le Brun}, {Maccagni}, {Tresse}, {Zamorani},
  {Bardelli}, {Bolzonella}, {Contini}, {Iovino}, {Lopez-Sanjuan}, {McCracken},
  {Pollo}, {Pozzetti}, {Scodeggio}, {Tasca}, {Vergani}, {Zanichelli}, \&
  {Zucca}}]{lefevre13}
{Le Fevre}, O., {Cassata}, P., {Cucciati}, O., {et~al.} 2013, ArXiv e-prints

\bibitem[{{Le F{\`e}vre} {et~al.}(2005){Le F{\`e}vre}, {Paltani}, {Arnouts},
  {Charlot}, {Foucaud}, {Ilbert}, {McCracken}, {Zamorani}, {Bottini},
  {Garilli}, {Le Brun}, {Maccagni}, {Picat}, {Scaramella}, {Scodeggio},
  {Tresse}, {Vettolani}, {Zanichelli}, {Adami}, {Bardelli}, {Bolzonella},
  {Cappi}, {Ciliegi}, {Contini}, {Franzetti}, {Gavignaud}, {Guzzo}, {Iovino},
  {Marano}, {Marinoni}, {Mazure}, {Meneux}, {Merighi}, {Pell{\`o}}, {Pollo},
  {Pozzetti}, {Radovich}, {Zucca}, {Arnaboldi}, {Bondi}, {Bongiorno},
  {Busarello}, {Gregorini}, {Lamareille}, {Mathez}, {Mellier}, {Merluzzi},
  {Ripepi}, \& {Rizzo}}]{lefevre05}
{Le F{\`e}vre}, O., {Paltani}, S., {Arnouts}, S., {et~al.} 2005, \nat, 437, 519

\bibitem[{{Le Fevre} {et~al.}(2014){Le Fevre}, {Tasca}, {Cassata}, {Garilli},
  {Le Brun}, {Maccagni}, {Pentericci}, {Thomas}, {Vanzella}, {Zamorani},
  {Zucca}, {Amorin}, {Bardelli}, {Capak}, {Cassara}, {Castellano}, {Cimatti},
  {Cuby}, {Cucciati}, {de la Torre}, {Durkalec}, {Fontana}, {Giavalisco},
  {Grazian}, {Hathi}, {Ilbert}, {Lemaux}, {Moreau}, {Paltani}, {Ribeiro},
  {Salvato}, {Schaerer}, {Scodeggio}, {Sommariva}, {Talia}, {Taniguchi},
  {Tresse}, {Vergani}, {Wang}, {Charlot}, {Contini}, {Fotopoulo},
  {Lopez-Sanjuan}, {Mellier}, \& {Scoville}}]{lefevre14}
{Le Fevre}, O., {Tasca}, L.~A.~M., {Cassata}, P., {et~al.} 2014, ArXiv e-prints

\bibitem[{{Leitherer} {et~al.}(2002){Leitherer}, {Li}, {Calzetti}, \&
  {Heckman}}]{leitherer02}
{Leitherer}, C., {Li}, I.-H., {Calzetti}, D., \& {Heckman}, T.~M. 2002, \apjs,
  140, 303

\bibitem[{{L{\'o}pez-Sanjuan} {et~al.}(2014){L{\'o}pez-Sanjuan}, {Cenarro},
  {Varela}, {Viironen}, {Molino}, {Ben{\'{\i}}tez}, {Arnalte-Mur}, {Ascaso},
  {D{\'{\i}}az-Garc{\'{\i}}a}, {Fern{\'a}ndez-Soto}, {Jim{\'e}nez-Teja},
  {M{\'a}rquez}, {Masegosa}, {Moles}, {Povi{\'c}}, {Aguerri}, {Alfaro},
  {Aparicio-Villegas}, {Broadhurst}, {Cabrera-Ca{\~n}o}, {Castander}, {Cepa},
  {Cervi{\~n}o}, {Crist{\'o}bal-Hornillos}, {Del Olmo}, {Gonz{\'a}lez Delgado},
  {Husillos}, {Infante}, {Mart{\'{\i}}nez}, {Perea}, {Prada}, \&
  {Quintana}}]{lopez-sanjuan14}
{L{\'o}pez-Sanjuan}, C., {Cenarro}, A.~J., {Varela}, J., {et~al.} 2014, ArXiv
  e-prints

\bibitem[{{Ly} {et~al.}(2011){Ly}, {Malkan}, {Hayashi}, {Motohara},
  {Kashikawa}, {Shimasaku}, {Nagao}, \& {Grady}}]{ly11}
{Ly}, C., {Malkan}, M.~A., {Hayashi}, M., {et~al.} 2011, \apj, 735, 91

\bibitem[{{Madau}(1995)}]{madau95}
{Madau}, P. 1995, \apj, 441, 18

\bibitem[{{Matute} {et~al.}(2012){Matute}, {M{\'a}rquez}, {Masegosa},
  {Husillos}, {del Olmo}, {Perea}, {Alfaro}, {Fern{\'a}ndez-Soto}, {Moles},
  {Aguerri}, {Aparicio-Villegas}, {Ben{\'{\i}}tez}, {Broadhurst},
  {Cabrera-Cano}, {Castander}, {Cepa}, {Cervi{\~n}o},
  {Crist{\'o}bal-Hornillos}, {Infante}, {Gonz{\'a}lez Delgado},
  {Mart{\'{\i}}nez}, {Molino}, {Prada}, \& {Quintana}}]{matute12}
{Matute}, I., {M{\'a}rquez}, I., {Masegosa}, J., {et~al.} 2012, \aap, 542, A20

\bibitem[{{Matute} {et~al.}(2013){Matute}, {Masegosa}, {M{\'a}rquez},
  {Fern{\'a}ndez-Soto}, {Husillos}, {del Olmo}, {Perea}, {Povi{\'c}}, {Ascaso},
  {Alfaro}, {Moles}, {Aguerri}, {Aparicio-Villegas}, {Ben{\'{\i}}tez},
  {Broadhurst}, {Cabrera-Cano}, {Castander}, {Cepa}, {Cervi{\~n}o},
  {Crist{\'o}bal-Hornillos}, {Infante}, {Gonz{\'a}lez Delgado},
  {Mart{\'{\i}}nez}, {Molino}, {Prada}, \& {Quintana}}]{matute13}
{Matute}, I., {Masegosa}, J., {M{\'a}rquez}, I., {et~al.} 2013, \aap, 557, A78

\bibitem[{{McLure} {et~al.}(2009){McLure}, {Cirasuolo}, {Dunlop}, {Foucaud}, \&
  {Almaini}}]{mclure09}
{McLure}, R.~J., {Cirasuolo}, M., {Dunlop}, J.~S., {Foucaud}, S., \& {Almaini},
  O. 2009, \mnras, 395, 2196

\bibitem[{{McLure} {et~al.}(2006){McLure}, {Cirasuolo}, {Dunlop}, {Sekiguchi},
  {Almaini}, {Foucaud}, {Simpson}, {Watson}, {Hirst}, {Page}, \&
  {Smail}}]{mclure06}
{McLure}, R.~J., {Cirasuolo}, M., {Dunlop}, J.~S., {et~al.} 2006, \mnras, 372,
  357

\bibitem[{{McLure} {et~al.}(2011){McLure}, {Dunlop}, {de Ravel}, {Cirasuolo},
  {Ellis}, {Schenker}, {Robertson}, {Koekemoer}, {Stark}, \&
  {Bowler}}]{mclure11}
{McLure}, R.~J., {Dunlop}, J.~S., {de Ravel}, L., {et~al.} 2011, \mnras, 418,
  2074

\bibitem[{{Moles} {et~al.}(2008){Moles}, {Ben{\'{\i}}tez}, {Aguerri}, {Alfaro},
  {Broadhurst}, {Cabrera-Ca{\~n}o}, {Castander}, {Cepa}, {Cervi{\~n}o},
  {Crist{\'o}bal-Hornillos}, {Fern{\'a}ndez-Soto}, {Gonz{\'a}lez Delgado},
  {Infante}, {M{\'a}rquez}, {Mart{\'{\i}}nez}, {Masegosa}, {del Olmo}, {Perea},
  {Prada}, {Quintana}, \& {S{\'a}nchez}}]{moles08}
{Moles}, M., {Ben{\'{\i}}tez}, N., {Aguerri}, J.~A.~L., {et~al.} 2008, \aj,
  136, 1325

\bibitem[{{Molino} {et~al.}(2014){Molino}, {Ben{\'{\i}}tez}, {Moles},
  {Fern{\'a}ndez-Soto}, {Crist{\'o}bal-Hornillos}, {Ascaso},
  {Jim{\'e}nez-Teja}, {Schoenell}, {Arnalte-Mur}, {Povi{\'c}}, {Coe},
  {L{\'o}pez-Sanjuan}, {D{\'{\i}}az-Garc{\'{\i}}a}, {Varela}, {Stefanon},
  {Cenarro}, {Matute}, {Masegosa}, {M{\'a}rquez}, {Perea}, {Del Olmo},
  {Husillos}, {Alfaro}, {Aparicio-Villegas}, {Cervi{\~n}o}, {Huertas-Company},
  {Aguerri}, {Broadhurst}, {Cabrera-Ca{\~n}o}, {Cepa}, {Gonz{\'a}lez},
  {Infante}, {Mart{\'{\i}}nez}, {Prada}, \& {Quintana}}]{molino14}
{Molino}, A., {Ben{\'{\i}}tez}, N., {Moles}, M., {et~al.} 2014, \mnras, 441,
  2891

\bibitem[{{Oke} \& {Gunn}(1983)}]{oke83}
{Oke}, J.~B. \& {Gunn}, J.~E. 1983, \apj, 266, 713

\bibitem[{{Oteo} {et~al.}(2013{\natexlab{a}}){Oteo}, {Bongiovanni}, {Cepa},
  {P{\'e}rez-Garc{\'{\i}}a}, {Ederoclite}, {S{\'a}nchez-Portal},
  {Pintos-Castro}, {P{\'e}rez-Mart{\'{\i}}nez}, {Polednikova}, {Aguerri},
  {Alfaro}, {Aparicio-Villegas}, {Ben{\'{\i}}tez}, {Broadhurst},
  {Cabrera-Ca{\~n}o}, {Castander}, {Cervi{\~n}o}, {Cristobal-Hornillos},
  {Fernandez-Soto}, {Gonzalez-Delgado}, {Husillos}, {Infante},
  {Mart{\'{\i}}nez}, {M{\'a}rquez}, {Masegosa}, {Matute}, {Moles}, {Molino},
  {Olmo}, {Perea}, {Povi{\'c}}, {Prada}, {Quintana}, \& {Viironen}}]{oteo13a}
{Oteo}, I., {Bongiovanni}, {\'A}., {Cepa}, J., {et~al.} 2013{\natexlab{a}},
  \mnras, 433, 2706

\bibitem[{{Oteo} {et~al.}(2013{\natexlab{b}}){Oteo}, {Magdis}, {Bongiovanni},
  {P{\'e}rez-Garc{\'{\i}}a}, {Cepa}, {Cedr{\'e}s}, {Ederoclite},
  {S{\'a}nchez-Portal}, {Aguerri}, {Alfaro}, {Altieri}, {Andreani},
  {Aparicio-Villegas}, {Aussel}, {Ben{\'{\i}}tez}, {Berta}, {Broadhurst},
  {Cabrera-Ca{\~n}o}, {Castander}, {Cervi{\~n}o}, {Cimatti},
  {Cristobal-Hornillos}, {Daddi}, {Elbaz}, {Fernandez-Soto}, {Schreiber},
  {Genzel}, {Gonzalez-Delgado}, {Husillos}, {Infante}, {Le Floc'h}, {Lutz},
  {Magnelli}, {Maiolino}, {M{\'a}rquez}, {Mart{\'{\i}}nez}, {Masegosa},
  {Matute}, {Moles}, {Molino}, {Olmo}, {Perea}, {P{\'e}rez-Mart{\'{\i}}nez},
  {Pintos-Castro}, {Poglitsch}, {Polednikova}, {Popesso}, {Povi{\'c}}, {Pozzi},
  {Prada}, {Quintana}, {Riguccini}, {Sturm}, {Tacconi}, {Valtchanov}, \&
  {Viironen}}]{oteo13b}
{Oteo}, I., {Magdis}, G., {Bongiovanni}, {\'A}., {et~al.} 2013{\natexlab{b}},
  \mnras, 435, 158

\bibitem[{{Ouchi} {et~al.}(2004{\natexlab{a}}){Ouchi}, {Shimasaku}, {Okamura},
  {Furusawa}, {Kashikawa}, {Ota}, {Doi}, {Hamabe}, {Kimura}, {Komiyama},
  {Miyazaki}, {Miyazaki}, {Nakata}, {Sekiguchi}, {Yagi}, \&
  {Yasuda}}]{ouchi04b}
{Ouchi}, M., {Shimasaku}, K., {Okamura}, S., {et~al.} 2004{\natexlab{a}}, \apj,
  611, 660

\bibitem[{{Ouchi} {et~al.}(2004{\natexlab{b}}){Ouchi}, {Shimasaku}, {Okamura},
  {Furusawa}, {Kashikawa}, {Ota}, {Doi}, {Hamabe}, {Kimura}, {Komiyama},
  {Miyazaki}, {Miyazaki}, {Nakata}, {Sekiguchi}, {Yagi}, \&
  {Yasuda}}]{ouchi04a}
{Ouchi}, M., {Shimasaku}, K., {Okamura}, S., {et~al.} 2004{\natexlab{b}}, \apj,
  611, 685

\bibitem[{{P{\^a}ris} {et~al.}(2014){P{\^a}ris}, {Petitjean}, {Aubourg},
  {Ross}, {Myers}, {Streblyanska}, {Bailey}, {Hall}, {Strauss}, {Anderson},
  {Bizyaev}, {Borde}, {Brinkmann}, {Bovy}, {Brandt}, {Brewington},
  {Brownstein}, {Cook}, {Ebelke}, {Fan}, {Filiz Ak}, {Finley}, {Font-Ribera},
  {Ge}, {Hamann}, {Ho}, {Jiang}, {Kinemuchi}, {Malanushenko}, {Malanushenko},
  {Marchante}, {McGreer}, {McMahon}, {Miralda-Escud{\'e}}, {Muna},
  {Noterdaeme}, {Oravetz}, {Palanque-Delabrouille}, {Pan}, {Perez-Fournon},
  {Pieri}, {Riffel}, {Schlegel}, {Schneider}, {Simmons}, {Viel}, {Weaver},
  {Wood-Vasey}, {Y{\`e}che}, \& {York}}]{paris14}
{P{\^a}ris}, I., {Petitjean}, P., {Aubourg}, {\'E}., {et~al.} 2014, \aap, 563,
  A54

\bibitem[{{P{\'e}rez-Gonz{\'a}lez} {et~al.}(2013){P{\'e}rez-Gonz{\'a}lez},
  {Cava}, {Barro}, {Villar}, {Cardiel}, {Ferreras},
  {Rodr{\'{\i}}guez-Espinosa}, {Alonso-Herrero}, {Balcells}, {Cenarro}, {Cepa},
  {Charlot}, {Cimatti}, {Conselice}, {Daddi}, {Donley}, {Elbaz}, {Espino},
  {Gallego}, {Gobat}, {Gonz{\'a}lez-Mart{\'{\i}}n}, {Guzm{\'a}n},
  {Hern{\'a}n-Caballero}, {Mu{\~n}oz-Tu{\~n}{\'o}n}, {Renzini},
  {Rodr{\'{\i}}guez-Zaur{\'{\i}}n}, {Tresse}, {Trujillo}, \&
  {Zamorano}}]{perez_gonzalez13}
{P{\'e}rez-Gonz{\'a}lez}, P.~G., {Cava}, A., {Barro}, G., {et~al.} 2013, \apj,
  762, 46

\bibitem[{{Reddy} {et~al.}(2006){Reddy}, {Steidel}, {Erb}, {Shapley}, \&
  {Pettini}}]{reddy06}
{Reddy}, N.~A., {Steidel}, C.~C., {Erb}, D.~K., {Shapley}, A.~E., \& {Pettini},
  M. 2006, \apj, 653, 1004

\bibitem[{{Reddy} {et~al.}(2008){Reddy}, {Steidel}, {Pettini}, {Adelberger},
  {Shapley}, {Erb}, \& {Dickinson}}]{reddy08}
{Reddy}, N.~A., {Steidel}, C.~C., {Pettini}, M., {et~al.} 2008, \apjs, 175, 48

\bibitem[{{Ross} {et~al.}(2013){Ross}, {McGreer}, {White}, {Richards}, {Myers},
  {Palanque-Delabrouille}, {Strauss}, {Anderson}, {Shen}, {Brandt},
  {Y{\`e}che}, {Swanson}, {Aubourg}, {Bailey}, {Bizyaev}, {Bovy}, {Brewington},
  {Brinkmann}, {DeGraf}, {Di Matteo}, {Ebelke}, {Fan}, {Ge}, {Malanushenko},
  {Malanushenko}, {Mandelbaum}, {Maraston}, {Muna}, {Oravetz}, {Pan},
  {P{\^a}ris}, {Petitjean}, {Schawinski}, {Schlegel}, {Schneider}, {Silverman},
  {Simmons}, {Snedden}, {Streblyanska}, {Suzuki}, {Weinberg}, \&
  {York}}]{ross13}
{Ross}, N.~P., {McGreer}, I.~D., {White}, M., {et~al.} 2013, \apj, 773, 14

\bibitem[{{Salvato} {et~al.}(2009){Salvato}, {Hasinger}, {Ilbert}, {Zamorani},
  {Brusa}, {Scoville}, {Rau}, {Capak}, {Arnouts}, {Aussel}, {Bolzonella},
  {Buongiorno}, {Cappelluti}, {Caputi}, {Civano}, {Cook}, {Elvis}, {Gilli},
  {Jahnke}, {Kartaltepe}, {Impey}, {Lamareille}, {Le Floc'h}, {Lilly},
  {Mainieri}, {McCarthy}, {McCracken}, {Mignoli}, {Mobasher}, {Murayama},
  {Sasaki}, {Sanders}, {Schiminovich}, {Shioya}, {Shopbell}, {Silverman},
  {Smol{\v c}i{\'c}}, {Surace}, {Taniguchi}, {Thompson}, {Trump}, {Urry}, \&
  {Zamojski}}]{salvato09}
{Salvato}, M., {Hasinger}, G., {Ilbert}, O., {et~al.} 2009, \apj, 690, 1250

\bibitem[{{Santini} {et~al.}(2009){Santini}, {Fontana}, {Grazian}, {Salimbeni},
  {Fiore}, {Fontanot}, {Boutsia}, {Castelllano}, {Cristiani}, {de Santis},
  {Gallozzi}, {Giallongo}, {Nonino}, {Menci}, {Paris}, {Pentericci}, \&
  {Vanzella}}]{santini09}
{Santini}, P., {Fontana}, A., {Grazian}, A., {et~al.} 2009, VizieR Online Data
  Catalog, 350, 40751

\bibitem[{{Scoville} {et~al.}(2007){Scoville}, {Abraham}, {Aussel}, {Barnes},
  {Benson}, {Blain}, {Calzetti}, {Comastri}, {Capak}, {Carilli}, {Carlstrom},
  {Carollo}, {Colbert}, {Daddi}, {Ellis}, {Elvis}, {Ewald}, {Fall},
  {Franceschini}, {Giavalisco}, {Green}, {Griffiths}, {Guzzo}, {Hasinger},
  {Impey}, {Kneib}, {Koda}, {Koekemoer}, {Lefevre}, {Lilly}, {Liu},
  {McCracken}, {Massey}, {Mellier}, {Miyazaki}, {Mobasher}, {Mould}, {Norman},
  {Refregier}, {Renzini}, {Rhodes}, {Rich}, {Sanders}, {Schiminovich},
  {Schinnerer}, {Scodeggio}, {Sheth}, {Shopbell}, {Taniguchi}, {Tyson}, {Urry},
  {Van Waerbeke}, {Vettolani}, {White}, \& {Yan}}]{scoville07}
{Scoville}, N., {Abraham}, R.~G., {Aussel}, H., {et~al.} 2007, \apjs, 172, 38

\bibitem[{{Shapley} {et~al.}(2003){Shapley}, {Steidel}, {Pettini}, \&
  {Adelberger}}]{shapley03}
{Shapley}, A.~E., {Steidel}, C.~C., {Pettini}, M., \& {Adelberger}, K.~L. 2003,
  \apj, 588, 65

\bibitem[{{Shim} {et~al.}(2007){Shim}, {Im}, {Choi}, {Yan}, \&
  {Storrie-Lombardi}}]{shim07}
{Shim}, H., {Im}, M., {Choi}, P., {Yan}, L., \& {Storrie-Lombardi}, L. 2007,
  \apj, 669, 749

\bibitem[{{Steidel} {et~al.}(2003){Steidel}, {Adelberger}, {Shapley},
  {Pettini}, {Dickinson}, \& {Giavalisco}}]{steidel03}
{Steidel}, C.~C., {Adelberger}, K.~L., {Shapley}, A.~E., {et~al.} 2003, \apj,
  592, 728

\bibitem[{{Steidel} {et~al.}(1996{\natexlab{a}}){Steidel}, {Giavalisco},
  {Dickinson}, \& {Adelberger}}]{steidel96a}
{Steidel}, C.~C., {Giavalisco}, M., {Dickinson}, M., \& {Adelberger}, K.~L.
  1996{\natexlab{a}}, \aj, 112, 352

\bibitem[{{Steidel} {et~al.}(1996{\natexlab{b}}){Steidel}, {Giavalisco},
  {Pettini}, {Dickinson}, \& {Adelberger}}]{steidel96b}
{Steidel}, C.~C., {Giavalisco}, M., {Pettini}, M., {Dickinson}, M., \&
  {Adelberger}, K.~L. 1996{\natexlab{b}}, \apjl, 462, L17

\bibitem[{{Steidel} \& {Hamilton}(1992)}]{steidel92}
{Steidel}, C.~C. \& {Hamilton}, D. 1992, \aj, 104, 941

\bibitem[{{Steidel} \& {Hamilton}(1993)}]{steidel93}
{Steidel}, C.~C. \& {Hamilton}, D. 1993, \aj, 105, 2017

\bibitem[{{Steidel} {et~al.}(2004){Steidel}, {Shapley}, {Pettini},
  {Adelberger}, {Erb}, {Reddy}, \& {Hunt}}]{steidel04}
{Steidel}, C.~C., {Shapley}, A.~E., {Pettini}, M., {et~al.} 2004, \apj, 604,
  534

\bibitem[{{Yoshida} {et~al.}(2006){Yoshida}, {Shimasaku}, {Kashikawa}, {Ouchi},
  {Okamura}, {Ajiki}, {Akiyama}, {Ando}, {Aoki}, {Doi}, {Furusawa},
  {Hayashino}, {Iwamuro}, {Iye}, {Karoji}, {Kobayashi}, {Kodaira}, {Kodama},
  {Komiyama}, {Malkan}, {Matsuda}, {Miyazaki}, {Mizumoto}, {Morokuma},
  {Motohara}, {Murayama}, {Nagao}, {Nariai}, {Ohta}, {Sasaki}, {Sato},
  {Sekiguchi}, {Shioya}, {Tamura}, {Taniguchi}, {Umemura}, {Yamada}, \&
  {Yasuda}}]{yoshida06}
{Yoshida}, M., {Shimasaku}, K., {Kashikawa}, N., {et~al.} 2006, \apj, 653, 988

\end{thebibliography}

\end{document}